\documentclass[aps, pre, notitlepage, superscriptaddress,twocolumn,
               a4paper, floatfix]{revtex4-1}

\usepackage[utf8]{inputenc}
\usepackage{epsfig}
\usepackage[T1]{fontenc}
\usepackage{t1enc}
\usepackage{graphicx}
\usepackage{amssymb}
\usepackage{amsmath}
\usepackage{relsize}
\usepackage[caption=false]{subfig}
\usepackage{color}

\usepackage[normalem]{ulem}

\usepackage{mathtools}
\def\multiset#1#2{\ensuremath{\left(\kern-.3em\left(\genfrac{}{}{0pt}{}{#1}{#2}\right)\kern-.3em\right)}}

\begin{document}

\title{Influence of the history force on inertial particle
advection: Gravitational effects and horizontal diffusion}

\author{Ksenia Guseva}
\email{ksenia.guseva@uni-oldenburg.de}
\affiliation{Theoretical Physics/Complex Systems, ICBM, University of Oldenburg, 26129 Oldenburg, Germany}

\author{Ulrike Feudel}
\email{ulrike.feudel@uni-oldenburg.de}
\affiliation{Theoretical Physics/Complex Systems, ICBM, University of Oldenburg, 26129 Oldenburg, Germany}
\affiliation{Institute for Physical Science and Technology, University of Maryland, College Park, MD 20742-2431, USA}

\author{Tam\'as T\'el}
\email{tel@general.elte.hu}
\affiliation{Institute for Theoretical Physics - HAS Research Group, E\"otv\"os University, P.O. Box 32, H-1518, Budapest, Hungary}


\begin{abstract}
We analyse the effect of the Basset history force on the sedimentation or rising
of inertial particles in a two-dimensional convection flow.  When memory effects
are neglected, the system exhibits rich dynamics, including periodic,
quasi-periodic and chaotic attractors. Here we show that when the full advection
dynamics is considered, including the history force, both the nature and the
number of attractors change, and a fractalization of their basins of attraction
appears. In particular, we show that the history force significantly weakens the
horizontal diffusion and changes the speed of sedimentation or rising. The
influence of the history force is dependent on the size of the advected
particles, being stronger for larger particles.
\end{abstract}

\maketitle

\section{Introduction}

The advection of finite size inertial particles in fluid flows is a subject of
research during the last decade (cf. \cite{cartwright_dynamics_2010} and
references therein). Theoretical studies consider the particle dynamics either
in simplified kinematic flows
\cite{nishikawa_advective_2001, nishikawa_finite-size_2002,zahnow_aggregation_2008} yielding chaotic
advection of particles, random \cite{bec_fractal_2003,zahnow_what_2009} or
turbulent flows \cite{ calzavarini_quantifying_2008}. Recent studies have
emphasized the importance of the phenomenon of preferential concentration,
observed in the advection of finite-size particles~\cite{bec_clustering_2005,
  calzavarini_quantifying_2008}. Inertial effects in such systems lead to the
appearance of attractors, which result in the tendency of particles to
accumulate in certain flow regions. Particular experiments have been conducted
to find these tendecies of particles to
cluster~\cite{fiabane_clustering_2012,gibert_where_2012}. The understanding of
the motion of finite size particles is crucial in many different disciplines of
science such as e.g.  the raindrop formation in cloud
microphysics~\cite{falkovich_acceleration_2002,drotos_chaotic_2011}, planet
formation in astrophysics~\cite{wilkinson_stokes_2008}, large scale advection of
biological species in oceanography~\cite{benczik_selective_2002,
  tel_chemical_2005, daewel_towards_2011}, waste water
treatment~\cite{zhang_modeling_2003}, in
engineering~\cite{crowe_multiphase_1998, michaelides_particles_2006} and is of a
particular relevance for sedimentation of marine
aggregates~\cite{logan_environmental_2012,zahnow_coagulation_2009,
  zahnow_particle-based_2011}.

The equations of motion for small spherical inertial particles were
formulated by Maxey and Riley~\cite{maxey_equation_1983} with
corrections by Auton et al.~\cite{auton_force_1988} and are
integro-differential in their full form. They contain an integral term
which accounts for the diffusion of vorticity around the particle
throughout its entire history.  This integral term is called the history
(or Basset) force~\cite{basset_treatise_2008}, and due to its difficult
computation it is often neglected. However, recent experimental and
numerical works~\cite{mordant_velocity_2000, candelier_effect_2004,
daitche_memory_2011,
  van_hinsberg_efficient_2011, vojir_effect_1994,
  daitche_advection_2012} have exposed the limitations of this
approximation.

In this work, we analyze the effect of the history force on particles in chaotic
advection in the presence of gravity. One of the basic phenomena in this class
of problems is sedimentation (rising) of heavy (light) particles, that might be
accompanied with vertical trapping and horizontal diffusion. The only previous
efforts to understand the importance of the history force in the presence of
gravity is due to Mordant and Pinton~\cite{mordant_velocity_2000} and to Lohse
and coworkers~\cite{toegel_viscosity_2006, garbin_history_2009} who also carried
out experiments. Their investigation was however concentrating on free
sedimentation, that is on the particle motion in a fluid at rest, and on bubble
dynamics in a standing wave respectively. Here we are interested in how fluid
motion effects sedimentation (and rising) in the presence of the history
force. To this end a paradigmatic convection
flow~\cite{chandrasekhar_hidrodynamics_1961, maxey_motion_1987} with a periodic
forcing is used. It has already been studied
extensively~\cite{nishikawa_advective_2001, nishikawa_finite-size_2002,
  zahnow_moving_2008, zahnow_aggregation_2008}, without the inclusion of the
history force.  From these works it is known that the advection of inertial
particles is characterized by several different regimes with periodic,
quasiperiodic or chaotic attractors, depending on the choice of
parameters. Starting from this knowledge, we systematically compare the
advection with and without memory, and observe drastic changes. In general, we
find that the presence of the history force alters the average speed of
sedimentation or rising. It tends to weaken the particles' horizontal diffusion,
and new attractors appear in the system.

The paper is organized as follows: In Sec.~\ref{sec:overview} we present an
overview of the equations of motion, of the approach used to compute the history
force and analyse the choice of parameters for our model. In
Sec.~\ref{sec:general} a general comparison is made between the dynamics with
and without memory effects, highlighting the overall effects of the history
force from the point of view of dynamical systems theory. In
Sec.~\ref{sec:cases}, we choose four representative cases, for both aerosol and
bubble particles, for a more detailed comparison.  We focus on the changes of
the basins of attractions, the vertical trapping of particles, the appearance of
new attractors, as well as the change in their characteristics of the ones present without the
history force. Finally, we move in Sec.~\ref{sec:transport} to the description
of the vertical and horizontal transport properties of the flow, and how they
are altered by the history force. Our final conclusions are given in
Sec.~\ref{sec:conclusion}.

\section{Overview}\label{sec:overview}
\subsection{Particle advection}

We analyze the advection of spherical, rigid particles with a small particle
Reynolds number in an incompressible and viscous fluid. The Lagrangian
trajectories of such particles are evaluated according to the Maxey-Riley
equation~\cite{maxey_equation_1983}, including the corrections by Auton and
coworkers~\cite{auton_force_1988}. In the full Maxey-Riley picture one describes
the dimensionless evolution of the particle position $\vec{x}(t)$ and velocity
$\vec{v}(t) = d\vec{x}/dt$ in a flow field $\vec{u}(\vec{x}, t)$ as

\begin{equation}
\frac{d\vec{v}}{dt} = A(\vec{u} -\vec{v}) + AW\vec{n} + \frac{3R}{2}\frac{D\vec{u}}{Dt} - \sqrt{\frac{9AR}{2\pi}}\int_0^t \frac{\frac{d(\vec{v} - \vec{u})}{d\tau}}{\sqrt{t-\tau}}d\tau,
\label{eq:MR}
\end{equation}
where $\vec{n}$ is the vertical unit vector pointing upwards. This form of the
equation holds when the particle is initialized at time zero with a velocity
coinciding with that of the fluid.  We have to distinguish the full derivative
along a fluid element and a particle trajectory, given by
\begin{equation*}
\frac{D}{Dt} = \frac{\partial }{\partial t} + \vec{u} \cdot \nabla \qquad \text{and} \qquad \frac{d }{dt} = \frac{\partial}{\partial t} + \vec{v} \cdot \nabla
\end{equation*}
respectively. The velocity of the particle changes due to the action of
different forces. The forces in Eq.~\ref{eq:MR} represent from left to right:
the Stokes drag, the gravity, the pressure force (which accounts for the force
felt by a fluid element together with the added mass force), and lastly the
Basset history force. The equation is written in dimensionless form, rescaled by
a characteristic velocity $U$ and a characteristic length scale $L$ of the
flow. The ratio
\begin{equation}
R =\frac{2\rho_f}{\rho_f + 2\rho_p}
\end{equation}
between the density of the particle $\rho_p$ and of the fluid $\rho_f$,
respectively, divides the particles into aerosols ($R < \frac{2}{3}$) and
bubbles ($R > \frac{2}{3}$).  Another dimensionless parameter in
Eq. \ref{eq:MR} is the inertial parameter
\begin{equation}
  A = \underbrace{ R \frac{9\nu}{2r_p^2}\frac{L}{U}}_{1/St},
\end{equation}
where $St$ can be called the Stokes number, which provides a dimensionless
relation between particle radius $r_p$ and kinematic viscosity $\nu$ of the
fluid. A value $A$ or $St$ of order unity corresponds to strong inertial
effects.  Additionally, parameter $W$ governs the vertical movement
\begin{equation}
{W} = w \left(\frac{3}{2}-\frac{1}{R}\right), 
\label{W}
\end{equation}
with
\begin{equation}
w = \frac{g L}{U^2}\frac{R}{A} 
\label{w}
\end{equation}
where $g$ is the gravitational constant.  Eq.~\ref{W} expresses the fact that
bubbles tend to rise and aerosols to sediment.

For convenience, we introduce a unit particle with a radius $r_1$. All particles
in the system are characterized by their relative size with respect to this unit
particle.  This relation is described by the size parameter $\alpha$, which
describes how many times the mass $m_{\alpha}$ of a given particle of density
$\rho_p$ and radius $r_{\alpha}$ is larger than the mass of the unit
particle. Hence, $m_{\alpha} = \alpha m_1$ and $r_p\equiv r_{\alpha}
=\alpha^{1/3} r_1$. This representation of the size turns out to be useful in
studies of aggregation processes (see e.g., \cite{zahnow_aggregation_2008}). The
inertial parameter scales therefore with the size parameter, $\alpha$, as
\begin{equation}
A_{\alpha} = \alpha^{-2/3} \underbrace{R\frac{9\nu}{2r_1^2}\frac{L}{U}}_{A_1}.
\label{A1}
\end{equation}
According to Eq.~\ref{w}, $w_{\alpha}$ is inversely proportional to
$A_{\alpha}$, and therefore
\begin{equation}
w_{\alpha} = \alpha^{2/3} w_1,
\label{w1}
\end{equation}
with $w_{1} = gLR/(U^2A_1)$.

\subsection{Flow}
The velocity field $\vec{u}(\vec{x},t)$ is chosen to be a paradigmatic model of a
convective cell flow, introduced in~\cite{chandrasekhar_hidrodynamics_1961,
  maxey_motion_1987} in its time independent form. It consists of
two-dimensional oscillating cellular vortices in the plane $(x_1',x_2')$,
represented in the time periodic form by the stream function
\begin{equation}
\psi' = \frac{UL}{\pi}\left[1+ k\sin \left(\omega \frac{U}{L} t'\right)\right] \sin\left(\pi \frac{x_1'}{L}\right)\sin\left(\pi \frac{x_2'}{L}\right),
\end{equation}
where $U$ and $L$ are the characteristic velocity and length, respectively, and
the remaining parameters are normally chosen as $k = 2.72$ and $\omega =
\pi$~\cite{nishikawa_finite-size_2002}. We work with the dimensionless
coordinates $x_i = x_i'/L$ for $i \in [1, 2]$, time $t = t' U/L$, and stream
function $\psi = \psi'/(UL)$. Each vortex is rotating in the opposite direction
of its four neighbours, and is subjected to a periodic forcing of dimensionless
period $T=2 \pi/\omega=2$.  The flow field follows from $\nabla \times
\vec{\Psi}$, where ${\vec{\Psi}} = (0, 0, \psi)$, leading to the dimensionless
velocity vector
\begin{equation}
\vec{u} = (1 + k\sin \omega t)\left(
                   \begin{array}{c}
                     \sin(\pi x_1)\cos(\pi x_2) \\
                     -\cos(\pi x_1)\sin(\pi x_2)
                   \end{array}
                   \right).
\end{equation}
The coordinate $x_2$ is considered to be the vertical one.

The flow is double periodic in both directions with a dimensionless spatial
period of $2$. We define the unit square containing a single vortex as a box,
and the two-by-two square of the four vortices as a cell (with two vortices in
the horizontal and two in the vertical directions).

At time zero the vortex in the left lower corner rotates counterclockwise.  Its
rotation speed increases up to time $t=1/2$ then it starts decreasing but
remains positive even at $t=1$. The speed changes sign only at
$t=\arcsin{(-1/k)}/\pi=1.12$, and a second time at $t=1.88$. The vortex rotates
thus in the negative direction for a period of length $0.76$ only, with a slower
average speed than in the positive direction (over a period of $1.24$).

The particle dynamics can be represented in different ways. When the particle
position is always shifted back to the cell $x_1, x_2 \in (0,2)$, we speak of
the double periodic representation. When the motion is monitored in the full
$x_1, x_2$ plane, we obtain the planar representation. Another distinction can
be made by following the particle in continuous time, or as a stroboscopic map
taken at integer periods of the (dimensionless) period $T=2$ of the flow.
 
\subsection{Choice of parameters}
The parameter that most strongly influences the dynamics is the particle size
$\alpha$. It is a remarkable feature of this model that the passive advection
problem ($k=0, A=\infty$) is integrable, and hence any kind of complex behavior
in this system is due to the inertia of the particles.

Recent studies of marine ecosystems have emphasized the importance of marine
snow, which are aggregates composed by organic (phytoplankton, bacteria) and
inorganic material~\cite{simon_microbial_2002}. Marine snow plays a central role
in the carbon cycle~\cite{mann_dynamics_2005,
  de_la_rocha_factors_2007,riley_relative_2012}, and its formation is mainly due to
physical coagulation, which requires understanding the particle-flow
interactions. The size of larger aggregates varies from $0.1$ to less then $1$
mm~\cite{de_la_rocha_factors_2007, bartholoma_suspended_2009}. Typical velocities in the
ocean's upper layer are strongly dependent on the wind and can reach up to $0.5$
m$/$s~\cite{oakey_dissipation_1982}. The dissipation of turbulent energy
$\epsilon$ was found to change from $10^{-10}$ to $10^{-3}$ m$^2/$s$^3$, which
sets the size of the smallest possible eddies to $\sim10^{-3}-10^{-2} $
m~\cite{mann_dynamics_2005}.
 
The choice of the parameter range for our studies is motivated by these
observations, and by the general argument that memory effects in the presence of
gravity are expected to be relevant in small scale flows of moderate velocity
fluctuations. As a typical case, let us consider particles of radius
$r_1=0.5$ mm in a flow of characteristic velocity $U=0.3$ m/s and of size
$L=0.1$ m.  With the kinematic viscosity of water $\nu \approx 10^{-6}$ m$^2/$s
we find both $A_1$ and $w_1$ to be of the order of unity. 


We therefore choose $A_1=3$, $w_1=1.6$, and investigate the size parameter range
$\alpha=(0.1,5)$.  This corresponds to changing the particle size from $0. 46
r_1 = 0.23$ mm to $1.7 r_1 = 0.85$ mm, which implies, in view of Eq.~\ref{A1}
and Eq.~\ref{w1}, changing $A$ and $w$ from $4.6 A_1$ to $0.34 A_1$ and from
$0.22 w_1$ to $2.9 w_1$, respectively. As density ratios we take $R=0.5$ and
$R=1$ as typical aerosol and bubble characteristics, respectively. Eq.~\ref{W}
indicates that the modulus of parameter $W$ is $w/2$ in both cases.
 
\subsection{On the history force and its computation}

The history force arises due to the diffusion of flow perturbation patterns from
the particle's surface through the fluid. It is given by the last term of
Eq.~\ref{eq:MR}:
\begin{equation}
F_h = -\sqrt{\frac{9AR}{2\pi}}\int_0^t \frac{\frac{d(\vec{v} - \vec{u})}{d\tau}}{\sqrt{t-\tau}}d\tau,
\end{equation}
and it is calculated for the whole path followed by the particle since its
starting point where $\vec{v}=\vec{u}$. The integral term Eq.~\ref{eq:MR} makes
the differential equation very difficult to solve. For a few cases this term can
be evaluated analytically via a Laplace transform~\cite{candelier_effect_2004},
however for the great majority of cases it has to be obtained via time-consuming
numerical computations. The construction of new algorithms to optimize this
demanding procedure is an ongoing challenge. We use the most recent approach
developed in~\cite{daitche_advection_2012}, and we shortly review in this
section its main steps. Defining a function $f(\tau)$ as
\begin{equation}
    f(\tau) = \frac{d(\vec{v} - \vec{u})}{d\tau},
\end{equation}
the scheme consists of breaking the integral into intervals
$[\tau_i,\tau_{i+1}]$ of length $h$ and Taylor expanding $f(\tau)$ inside each
of these intervals.  After simple algebraic manipulations the first order
approximation is given by
\begin{equation}
\begin{array}{l}
\displaystyle \int_0^t \frac{f(\tau)}{\sqrt{t-\tau}}d\tau  = 2f(0)\sqrt{t} +  O(h^2)\sqrt{t}\\
\displaystyle + \frac{4}{3}\sum^{N-1}_{i=0} \frac{f(\tau_{i+1}) - f(\tau_{i})}{h}\left((t-\tau_i)^{\frac{3}{2}} -(t-\tau_{i+1})^{\frac{3}{2}}\right).
\end{array}
\end{equation}
The higher order terms can also be evaluated without difficulties, for more
details on the numerical implementation see~\cite{daitche_advection_2012}. The
results of the paper are obtained with the second order scheme but we checked
that a higher accuracy does not change the statistical properties. For the
solution of the ordinary differential equation we use the Runge-Kutta method
with a fixed time step $h = 0.01$, which is the same value as taken for the
discretization of the integral term.

\section{General analysis}\label{sec:general}

Our main focus is to compare the general aspects of the advective dynamics
computed through the full form of the Maxey-Riley equations (Eq.~\ref{eq:MR})
with the approximation which neglects the history force. We consider
Eq.~\ref{eq:MR} as a dynamical system and are interested in the properties of
the system for different particle sizes $\alpha$. The diversity of possible
behaviors was already described in ~\cite{nishikawa_finite-size_2002,
  zahnow_moving_2008, zahnow_aggregation_2008} by neglecting the history force.

\subsection{Bifurcation diagrams}\label{sec:bifurcation}

\begin{figure}[h]
  \begin{center}
    \includegraphics[width=.23\textwidth]{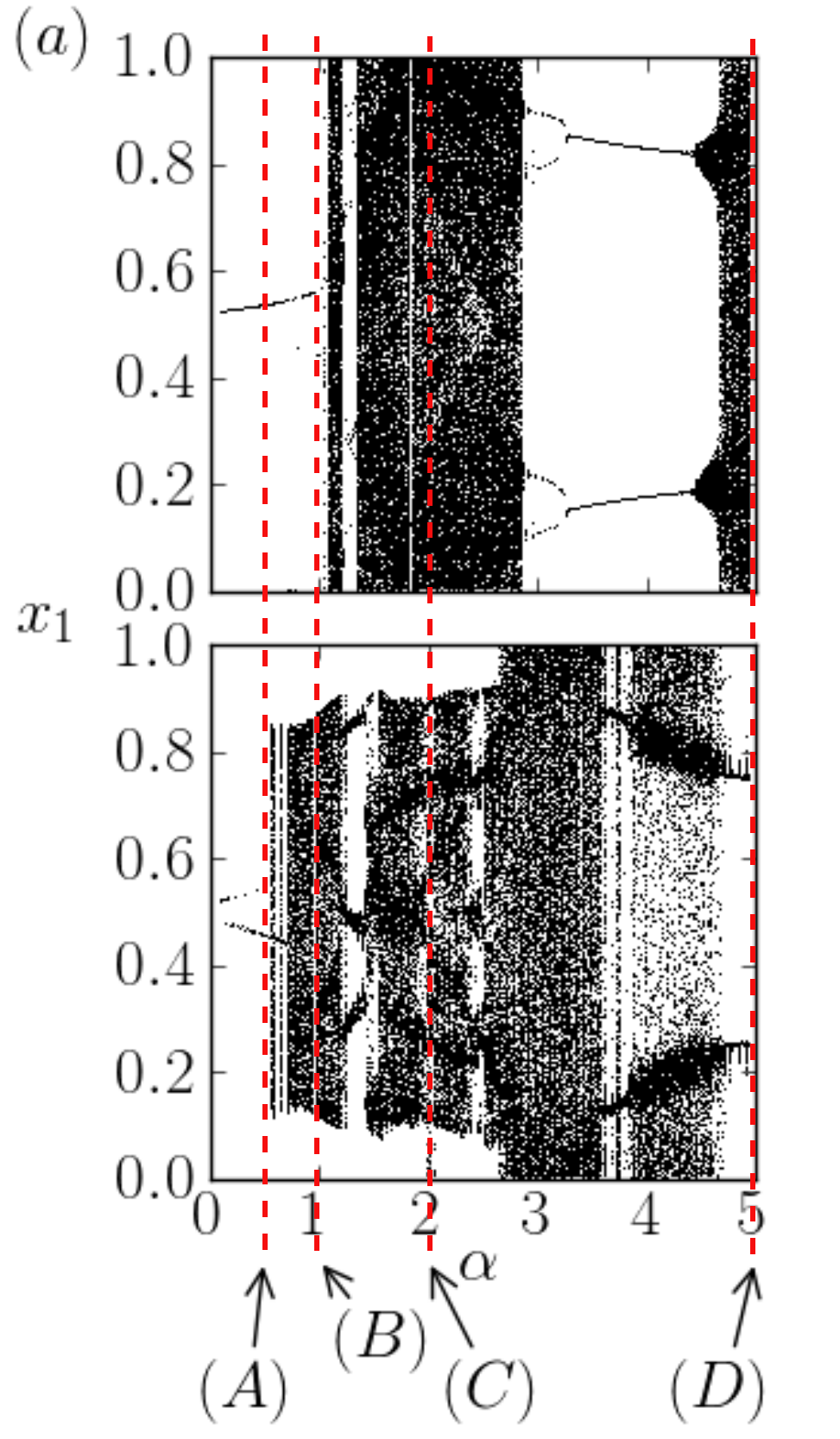}
    \includegraphics[width=.23\textwidth]{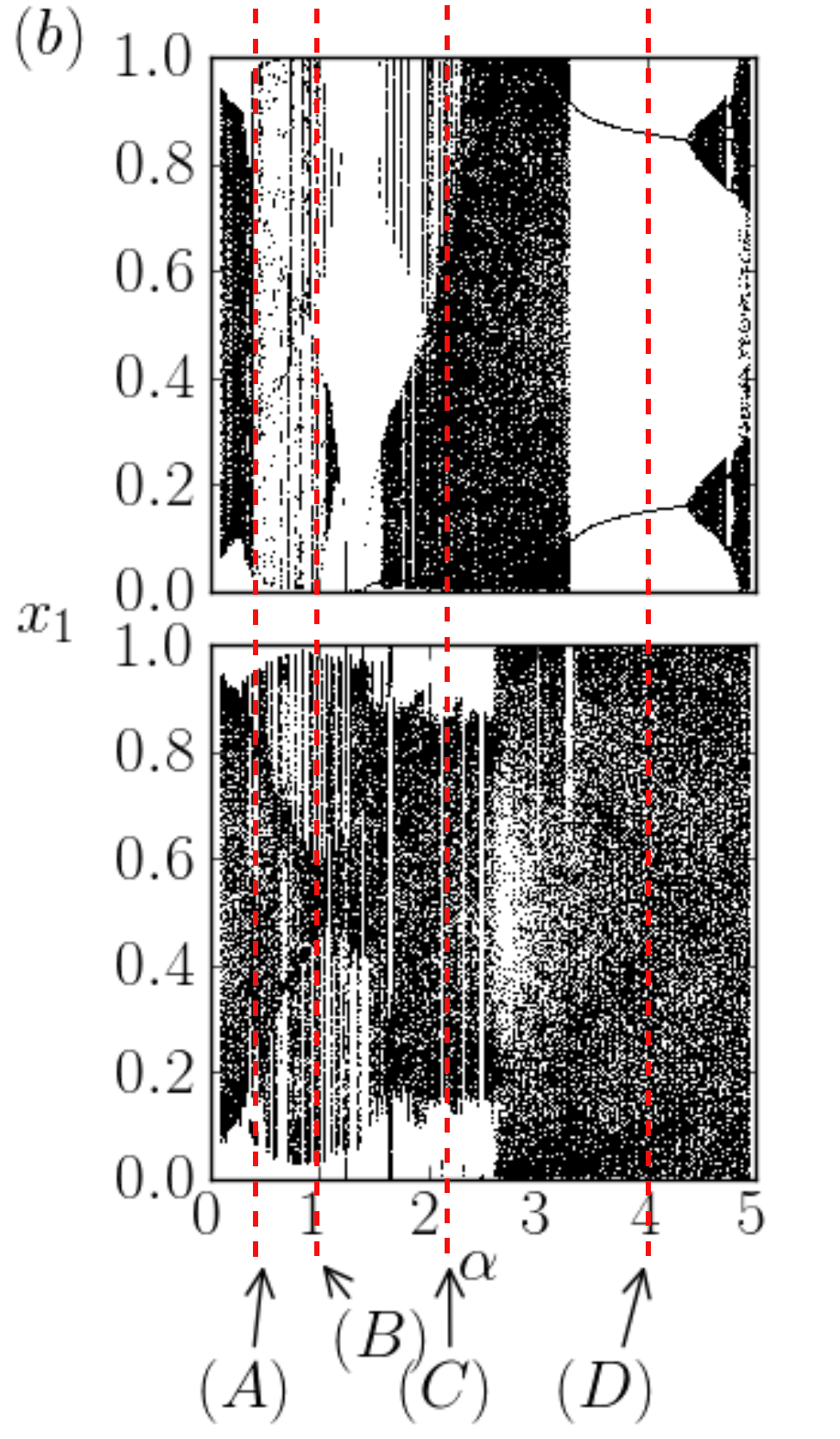}
    \end{center}
  \caption{Bifurcation diagrams: (a) Bubbles ($R = 1.0$); (b) Aerosols ($R =
    0.5$), without (top) and with (bottom) history force.In order to fit the
    diagram into the unit interval values of $2 > x_1 > 1$ are plotted as
    $1-x_1$. The arrows point to the $\alpha$ values (A, B, C, D) chosen as case
    studies to illustrate the strong changes in the dynamics with inclusion of
    memory in Sec.~\ref{sec:cases}.}\label{bifurcation}
\end{figure}

We start by analysing general aspects. For this purpose, we compare the
bifurcation diagrams with and without history force in the size parameter range
chosen. We initialize a particle in a {randomly chosen} position within a cell
with the flow velocity at that point {as its initial velocity}. Afterwards, the
particle is left to evolve according to Eq.~\ref{eq:MR} with and without the
history force. We discard long transients ($100 T$) and project the trajectory
into the double periodic representation. A stroboscopic sequence of the
$x_1$-coordinate of the particle is plotted as black dots in
Fig.~\ref{bifurcation}, for a time interval of $300T$ and $600T$ without and
with the history force, respectively. For both bubbles and aerosols, the
attractors are reached within the given time interval when memory effects are
not taken into account. It is important to emphasize that with memory much
longer times are necessary to reach the attractor (as will be clear in
Sec.~\ref{transient}). Therefore, what we see in the bifurcation diagrams in the
bottom of Fig.~\ref{bifurcation} are only rough estimates of the attractors, due
to the limited time span used. Nevertheless, the large differences between both
diagrams are quite evident in Fig.~\ref{bifurcation}.

\subsection{Slow transient dynamics with memory}\label{transient}

The basic difference in the transient behaviour with and without the history
force is illustrated by a case of bubbles with 4 coexisting point attractors in
the stroboscopic map. A detailed study of the dynamics will be given within case
A of Sec.~\ref{sec:cases}, here we concentrate only on the time needed to reach
the attractor (depicted in Fig.~\ref{fig:0.5traj}a).
\begin{figure}[h!]
  \begin{center}
    \subfloat[]{\includegraphics[width=.25\textwidth]{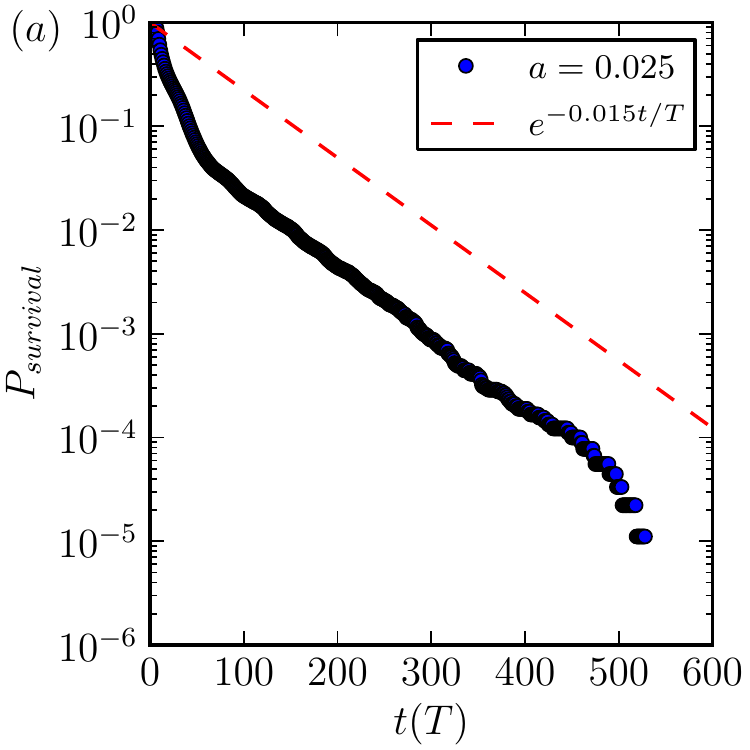}}
    \subfloat[]{\includegraphics[width=.25\textwidth]{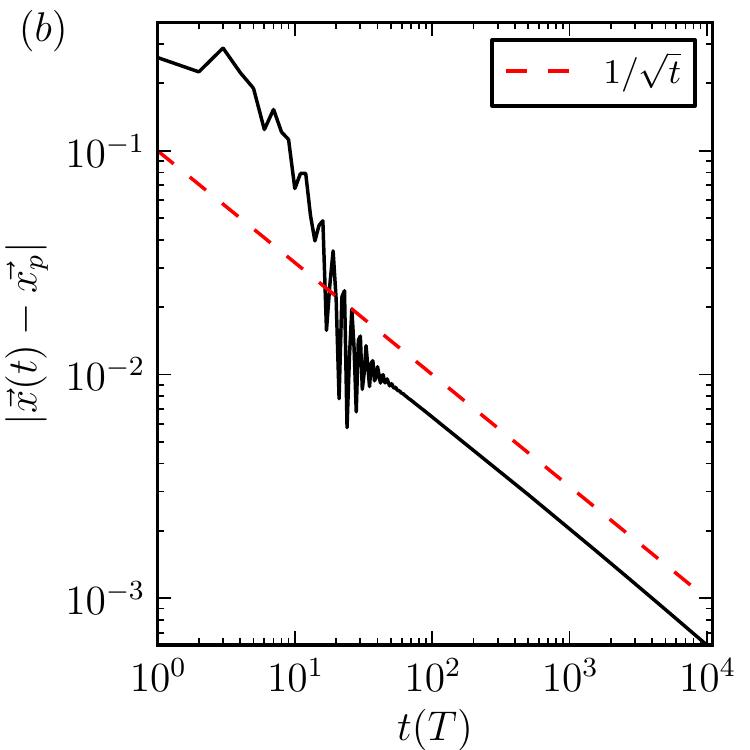}}
    \end{center}
  \caption{(a) Survival probability outside the four circles of radius $a = 0.025$
    around the attractor points as a function of time without history
    force obtained for an ensemble of particles. (b) The distance to the
    attractor as a function of time for a single particle initialized in
    the basin of attraction, with history force. The used parameters are
    $R = 1.0$, $\alpha=0.5$, and time is measured in units of the period
    $T$ of the flow.}\label{fig:escape}
\end{figure}

For the case without memory, it is numerically feasible to work with an ensemble
of particles and to estimate the survival probabilities in a region outside the
attractors. The particles are initialized homogeneously in space, and their
positions are recorded at integer multiples of $T$. Around each attractor point
(there are altogether 4 coexisting attractors) a circular disk of radius $a$ is
prescribed and we determine the number of particles outside these disks. In
Fig.~\ref{fig:escape}a we see that the normalized number of particles staying
outside the disks, the survival probability $P_{\text{survival}}$, decreases
exponentially in time. This points to the existence of a chaotic
saddle~\cite{lai_transient_2011} which is discussed in more detail in
Sec.~\ref{sec:cases}.  The particles outside the disk spend a long time on the
chaotic saddle where they move upwards from one cell to the next. The escape
rate from the chaotic saddle is found to be $\kappa=0.015/T$, out of which the
average lifetime can be estimated via $1/\kappa$ as about $66 T=132$ time
units. This implies that the number of survivors decreases by a factor of $1000$
after $500 T$ (Fig.~\ref{fig:escape}a).

When taking into account the memory force, we also find four point attractors
but they are reached extremely slowly. Due to the expensive calculation of the
integral term up to $10^4$ periods, we have to restrict our calculation to the
trajectory of a single particle. We initialize this particle within the basin of
attraction of one of the attractors and monitor the distance
$|\vec{x}(t)-\vec{x}_p|$ of this particle to the attractor point $\vec{x}_p$ in
the stroboscopic map (Fig.~\ref{fig:escape}b). We find that the long term
behavior is not exponential, but rather a power law decay proportional to
$t^{-1/2}$, as also observed in~\cite{daitche_memory_2011}. This diffusive
behavior is a consequence of the history kernel in Eq.~\ref{eq:MR}. The decay is
obviously much slower than the one without memory. Here we chose the initial
condition of the particle such that it does not rise but is trapped in the initial
box. We see that even if trapping is immediate, the distance to the attractor
decreases by only a factor of $10$ over the time interval of
$500T$. Qualitatively, the existence of a periodic attractor can be explained by
the observation that after a long time the trajectory is close to but not yet on
the attractor, the memory of the early approach to the attractor decays,
and the dynamics remembers only the motion in the close vicinity of the
attractor, and a convergence becomes thus
possible~\cite{daitche_memory_2011}. This mechanism also explains the long time
needed for convergence.  It is this slow power-law decay --- coupled with the
demanding numerical computations --- that makes a precise numerical
determination of all the attractors in the bifurcation diagram in the presence
of the history force impossible. Due to this technical difficulty, in this work
we shall follow trajectories or ensembles of trajectories for a sufficiently
long time, plot the positions of the trajectories at this instant, and call this
an (approximate) attractor.

\subsection{Relative magnitude of the history force}

\begin{figure}[h]
  \begin{center}
    \subfloat[]{\includegraphics[width=.25\textwidth]{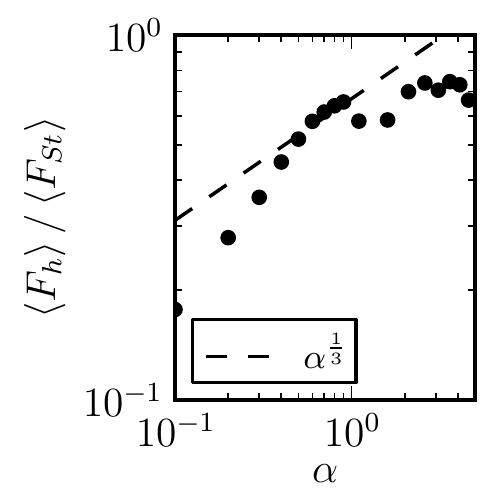}}
    \subfloat[]{\includegraphics[width=.25\textwidth]{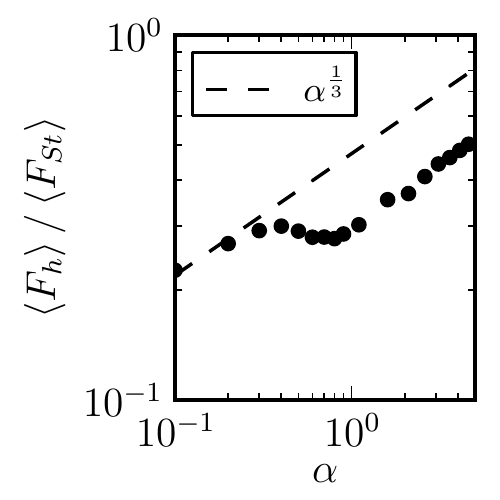}}\\
  \end{center}
  \caption{Average of the magnitude of the history force compared to the Stokes
    drag as a function of particle size, $\alpha$, for (a) $R=1.0$ and (b)
    $R=0.5$. For each parameter an ensemble of 10 random initial conditions in a
    cell are taken. The forces are measured and averaged over a time range of
    $500T$, and afterwards an ensemble average is also determined. The dashed
    lines represent a $\alpha^{1/3}$ curve, for
    comparison.}\label{fig:forcas_dir}
\end{figure}

As another general aspect, we see from the last term of Eq.~\ref{eq:MR} that the
history force is proportional to ${A}^{1/2}$, therefore its influence is
expected to be relatively more dominant for larger particles (smaller A). The
ratio of the history $F_h(t)$ and the drag force $F_{St}(t)$ (at fixed
densities) can be estimated from Eq.~\ref{eq:MR} as $A^{1/2}/A = A^{-1/2}$,
although the exact value will also depend on the velocities of flow and
particle, and their changes along the trajectory. In view of Eq.~\ref{A1} this
implies a proportionality to $\alpha^{1/3}$. This tendency is in harmony with
simulation results presented in Fig.~\ref{fig:forcas_dir}a and b, where the
average ratio $\left<F_h(t)\left>/\right<F_{St}(t)\right>$ is plotted as a
function of particle size $\alpha$. The deviations from the simple power law
scaling are due to the fact that the integral in Eq.~\ref{A1} and $v-u$ are not
always of the same order in modulus. Note that the average of the history force
can grow up to $80\%$ ($50\%$) of the Stokes drag for bubbles (aerosols).  All
this indicates that the history force must not be neglected.

Although the dynamics for bubbles and aerosols are very distinct, the
history force can produce similar effects in both cases. We illustrate
this by analyzing in detail four specific cases, shown in Fig,~\ref{bifurcation}.

\section{Case studies}\label{sec:cases}

The cases are chosen in order to illustrate a fractalization of the basin
boundaries and a deformation of the attractor (A), changes in transport
characteristics (in trapping) (B), appearance of new attractors (C), and changes
in the type of attractors (D). In the following subsections we will describe in
detail each of these cases with one example for bubbles and one for aerosols.

\subsection{Fractalization of the basins of attraction, deformation of the attractors}

\begin{figure}[h]
  \begin{center}
    \includegraphics[width=.5\textwidth]{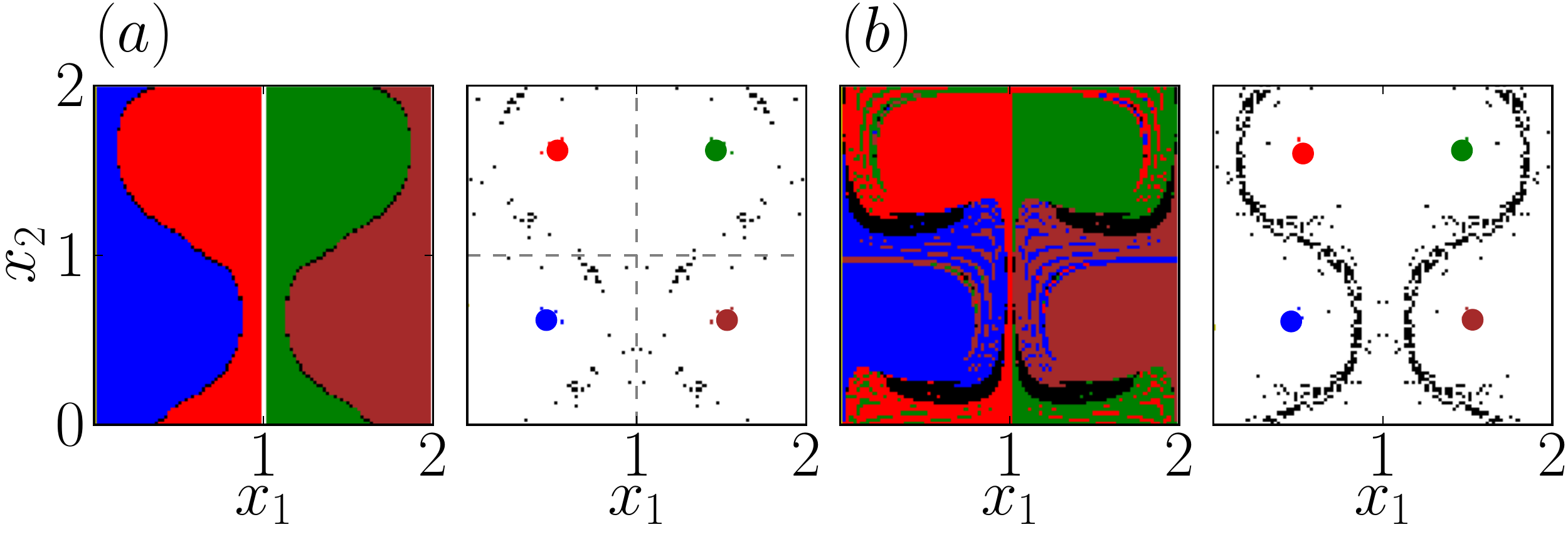}\\
    \includegraphics[width=.5\textwidth]{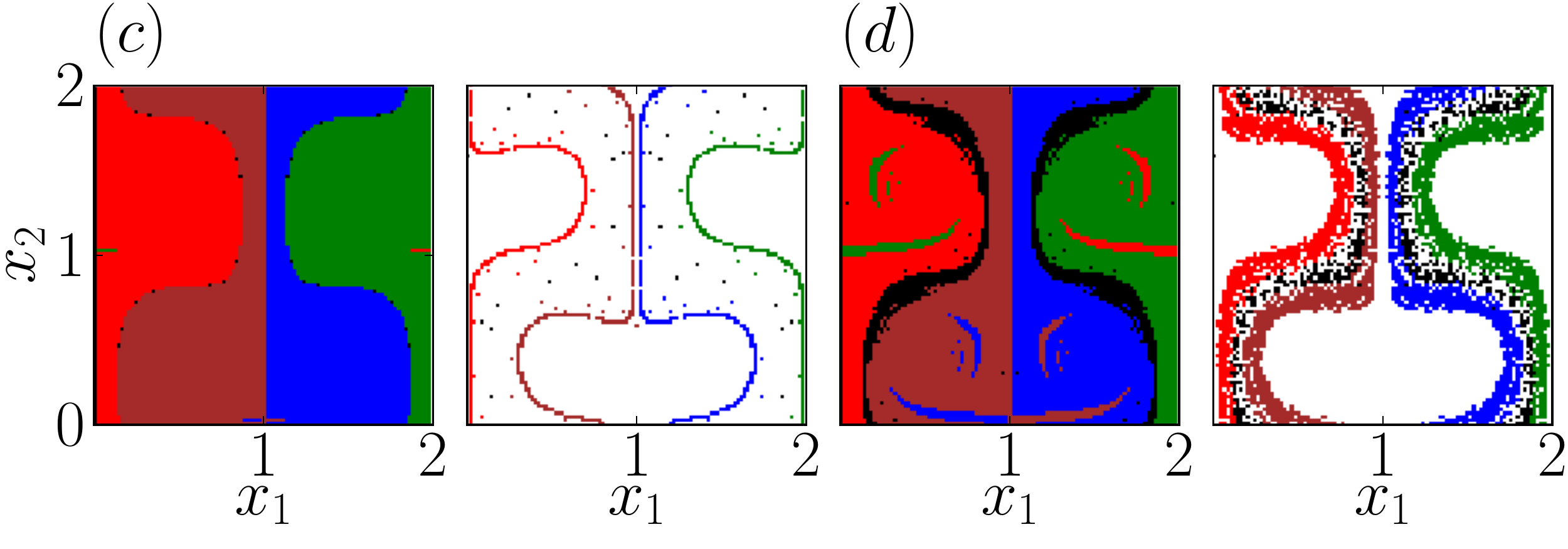}
    \end{center}
  \caption{Comparisons of basins and attractors for $\alpha=0.5$ (a, c) without
    memory, (b, d) with memory, for (a, b) bubbles ($R = 1.0$) and (c, d)
    aerosols ($R = 0.5$).  On the left and right of each panel are the basins of
    attraction and the four corresponding attractors (in red, blue, green and
    brown) obtained for $10^4$ particles at $t=120T$ in stroboscopic map in
    a double periodic representation.  For better visibility we mark in the
    right figure of panel (a) the boxes (unit squares) within a cell. Points of
    the chaotic transients, that is particles which have not yet reached the
    attractors after $120T$, appear in black.} \label{fig:basin_0.5}
\end{figure}

We start our case study with very small particles ($\alpha = 0.5$) where the
effects of the history force are relatively mild. One can read of
Fig.~\ref{fig:forcas_dir} that the history force corresponds here in magnitude
to $\sim 50\%$ of the Stokes drag for bubbles and almost $30\% $ of the Stokes
drag for aerosols. In both cases, bubbles and aerosols, we find four coexisting
periodic, limit cycle attractors, each of which having their own basin of
attraction, i.e. set of initial conditions that converge to that attractor.
Although, the dynamics with and without history force appear to be very similar
in this case, there are already clear differences. The most striking of them is
that the basins of attraction change their topology. While the basin boundaries
appear to be smooth without history force, they acquire a pronounced fractal
character when the history force is included (see
Fig.~\ref{fig:basin_0.5}). These changes are of course consequences of changes
in the particle dynamics. As a further illustration of this, we performed
simulations with an ensemble containing a large number of particles ($N =
10^4$). This particle ensemble is monitored over $120 T$. A longer interval is
not possible to choose due to the numerical cost of recording the history force
with small time step for this number of particles.

Bubbles are attracted to the vicinity of the vortex center, mainly because of
the effect of the pressure force. Once there, they move on a closed periodic
trajectory with the period $T$ of the flow. Therefore, in the double periodic
representation in the stroboscopic map, these limit cycles appear as four point
attractors: one in the neighborhood of the center of each box
(Fig.~\ref{fig:basin_0.5}a, b). 

Aerosols, on the other hand, switch quasi-periodically from the vicinity of box
centers to box boundaries, due to the time dependence of the flow, while moving
from one cell to the next one downstream. In the process of sedimentation, they
become confined horizontally to a region equivalent to one box size. We observe
four quasiperiodic attractors. In a double periodic representation of the
memoryless case they correspond to thin curves
(Fig.~\ref{fig:basin_0.5}c). 
In the presence of memory we find four color bands meandering about the vortices
(Fig.~\ref{fig:basin_0.5}d). The fattened appearance of these attractors is due
to the presence of the long transients with power-law decay discussed in the
previous section (Sec.~\ref{sec:general}B).

Although the history force does modify the location and shape of the attractors
for $\alpha = 0.5$, their main character remains unchanged. One of such
important characteristics for bubbles is the average rising height before
becoming trapped by a vortex. The average rising speed of a particle close to
the chaotic saddle is found to be $0.3$ box length per time unit without
memory. Multiplying this with the average lifetime of $66T$, obtained in
Sec.~\ref{sec:general}B (see Fig.~\ref{fig:escape}a), we find that bubbles
typically rise $40$ length units (about $20$ cells) before becoming
trapped. With memory, the great majority of initial conditions leads to a
dynamics where particles get trapped by the vortices within the first
$100T$. Although they take significantly longer times to reach the attractor
within a single cell (see. Fig.~\ref{fig:escape}b), the average trapping time to
a single box is similar to the memoryless case.

\begin{figure}[h]
  \includegraphics[width=.5\textwidth]{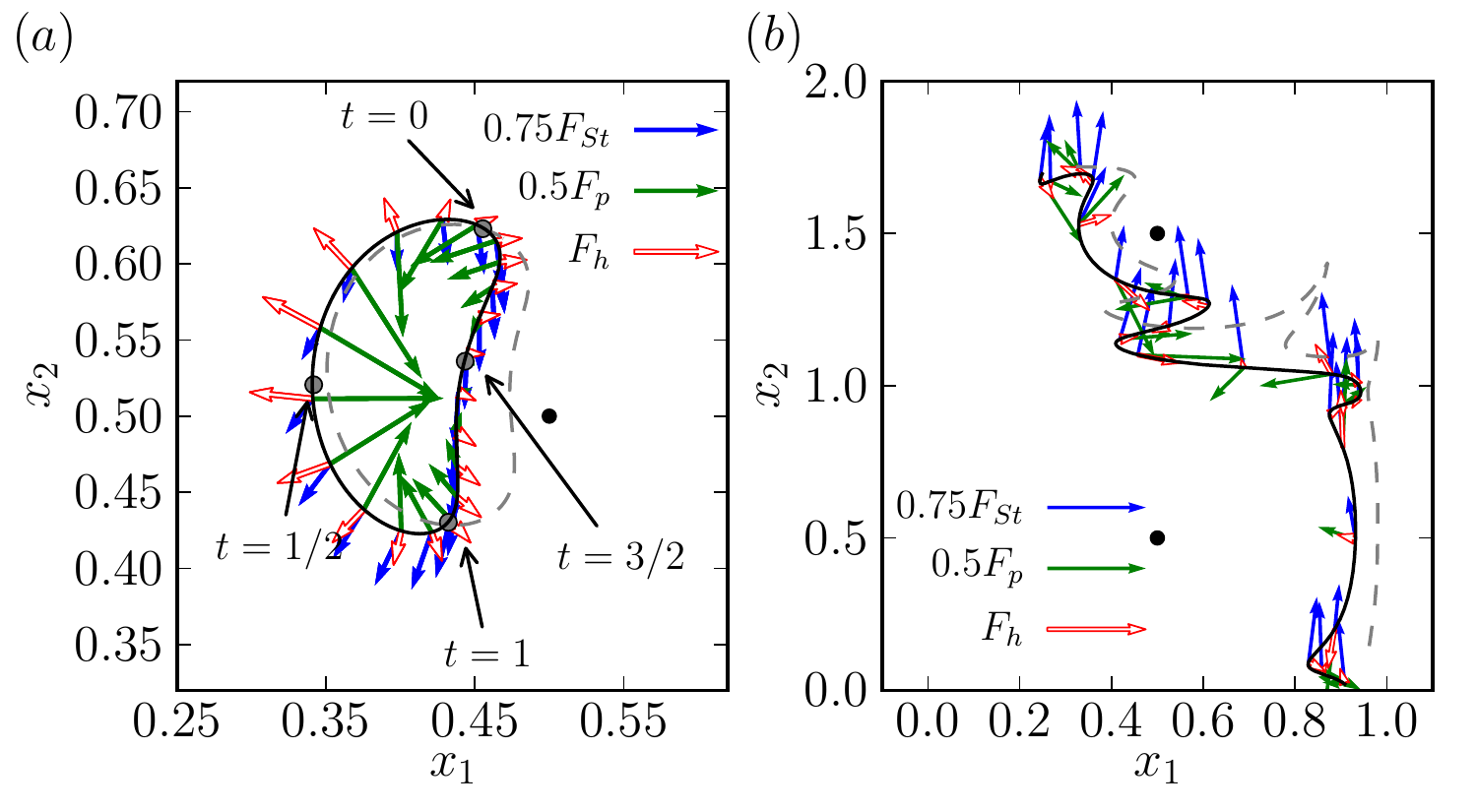}
  \caption{Force vectors along the attractor ($80T$ transients discarded) in a
    continuous-time planar representation.  The history force and (for better
    visibility) three quarters of the Stokes drag and one half of the pressure
    force are shown at a few instances within a time interval of about $8T$.
    (a) $R=1.0$. Time instants corresponding to quarter periods are indicated
    along the limit cycle attractor. (b) $R=0.5$. The trajectory with history
    force is plotted over a shorter interval of about $ 6.5T$. In both panels
    black dots represent the center of a vortex and the dashed gray line is the
    attractor without the history force. Since the overall vertical difference
    is approximately the same in panel (b), the difference in times implies an
    increased sedimentation speed with memory.}\label{fig:0.5traj}
\end{figure}

In spite of similarities, there are, however, considerable local changes of the
dynamics which we illustrate by plotting the attractors and the force vectors
along pieces of them {in a continuous time representation}
(Fig.~\ref{fig:0.5traj}). For bubbles, we see that the history force (in red)
changes direction roughly with the change of the velocity of the flow. We
illustrate this with the attractor in the left lower box of the cell
(Fig.~\ref{fig:0.5traj}a) where we also mark the time instants $t=0,1/2,1,3/2$
along the limit cycles. The vectors of the history force change direction indeed
at $t \sim 0$ and $t\sim1$.  When the particle is on the outer branch of the
attractor, the vortex motion is counterclockwise, and pushes the bubble
downwards. Along the inner branch (in the time interval $(1.12,1.88)$) the
vortex is rotating clockwise and the fluid motion helps the rising of the
bubble.  The history force is seen to counteract the pressure force (in
green). The resulting attractor is shifted further away from the vortex center
then the one without memory.

For aerosols, the attractors contain regions of large curvatures which
correspond to instants when the vortex rotations change sign.  The history force
drives particles on average away from the box center, and away from box
boundaries when they eventually approach them. The trajectory contains a lower
number of sharp curvatures, and the attractor is slightly shifted horizontally,
as can be seen in Fig.~\ref{fig:0.5traj}b. The fact that the trajectories are
more smooth implies that particles sediment slightly faster. In this particular example,
the average vertical speed changes from $\sim -0.125$ box length per time unit
without memory to $\sim -0.15$ box length per time unit with memory.

\subsection{Vertical leaking}
Next we investigate the behaviour of particles with $\alpha = 1.0$. Their
dynamics, without memory, consist of vertical motion until these particles get
captured in one of four possible limit cycle attractors
(Fig.~\ref{fig:traj1,0}a and b), for both bubbles and aerosols. For bubbles, the
attractors are periodic with period $T$. Aerosols have more complicated limit
cycles of period $12T$.
In the presence of the history force, we observe that particles cannot get
trapped vertically and move along extended quasiperiodic attractors
(Fig.~\ref{fig:traj1,0}c and d). Hence, the history force leads to vertical
leaking in this case. The history force is strong enough to essentially modify
the original periodic attractors into quasiperiodic ones. Bubbles are observed
to converge to two symmetric quasiperiodic attractors, while for aerosols the
number of attractors remains four (they appear in symmetric pairs).  In this
latter case, the sedimentation velocities on the different attractors do not
differ. What changes is the time to reach the attractor for different initial
conditions (this is reflected by the different locations of the two pairs of
attractor pieces plotted after the same time in Fig.~\ref{fig:traj1,0}d). The rising velocity of the bubble
(Fig.~\ref{fig:traj1,0}c) is found to be $\sim 0.31$ box length/ time unit. The
sedimentation velocity of the aerosols with memory is $\sim -0.33$ box length /
time unit (Fig.~\ref{fig:traj1,0}d).

\begin{figure}[h]
  \includegraphics[width=.5\textwidth]{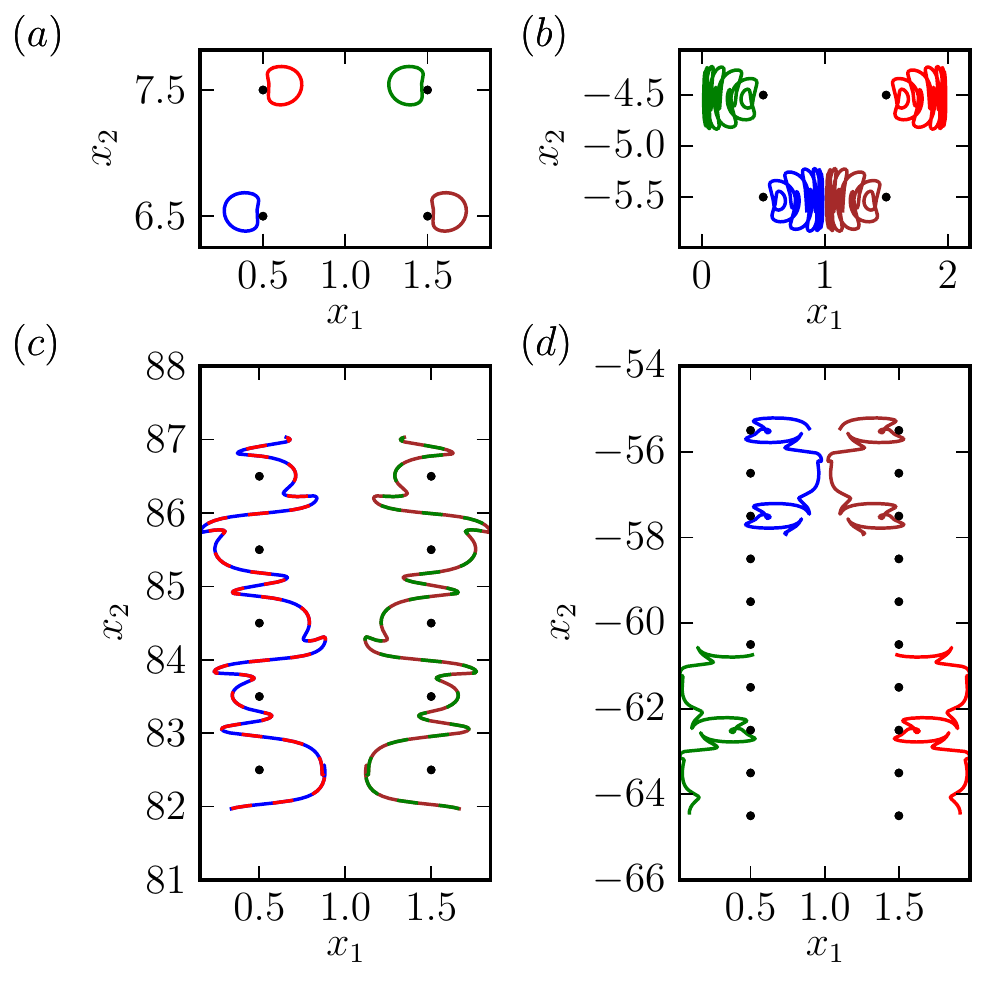}
  \caption{Comparison of attractors pieces for $\alpha=1$ in continuous time in
    the planar representation (trajectories are followed over a period of $20T$,
    after discarding transients of length $80T$). (a) bubble and (b) aerosol
    dynamics without history force. (c) bubble and (d) aerosol dynamics with
    memory effects. The black dots represent the center of
    vortices.}\label{fig:traj1,0}
\end{figure}

In summary, the effects of the history force in the case of $\alpha = 1$ are
that trapping is no longer possible, and the attractors change from periodic to
quasiperiodic, and vertical leaking occur. Additionally, the number of
attractors changes for bubbles. 


\subsection{Appearance of new attractors, and horizontal localization}

\begin{figure}[h]
  \begin{center}
    \includegraphics[width=.45\textwidth]{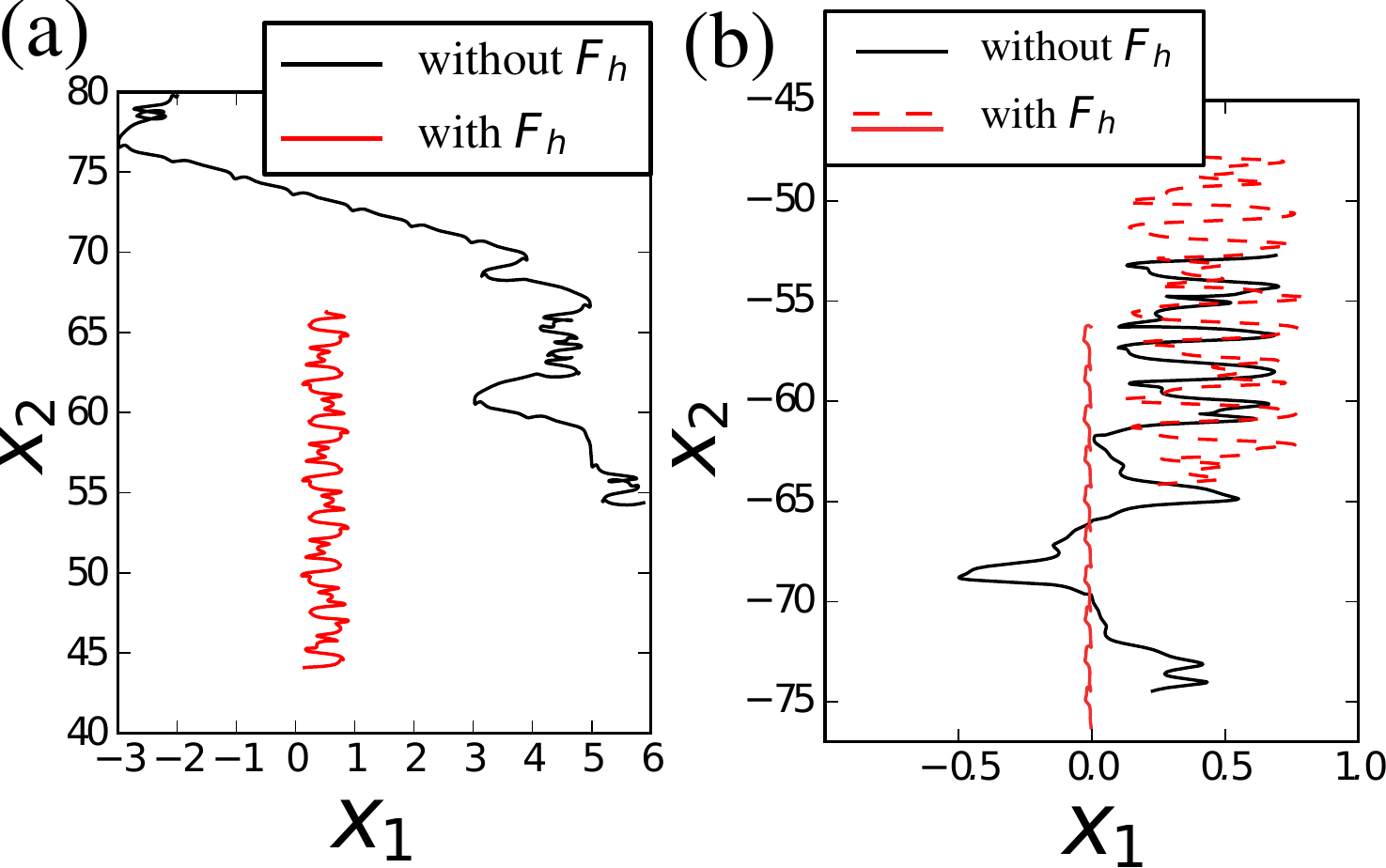}
    \end{center}
  \caption{Continuous time trajectories over $10T$ for (a) bubbles and (b)
    aerosols. In both simulations a transient of $50T$ is discarded.  Without
    memory a trajectory on the chaotic attractor is shown in (a) which strongly
    spreads horizontally. With memory a confined periodic attractor is traced
    out (the red attractor of Fig.~\ref{fig:bub_2,0}c). In (b) part of one
    of the chaotic attractors is shown without memory (the red attractor in
    Fig.~\ref{fig:basin_2,2}b, which communicates with its neighboring
    cell. With memory we display two of the six quasi periodic attractors (I'
    and II' of Fig.~\ref{fig:basin_2,2}d)}\label{fig:diff_traj}
\end{figure}

We illustrate that new types of attractors appear due to memory effects for
aerosols and bubbles of size $\alpha \sim 2$. A related feature is that the
history force prevents the particles to cross the vertical boundaries of the
cell. This confinement of the dynamics breaks off the horizontal diffusion
present without memory (Fig.~\ref{fig:diff_traj}), and results in the emergence
of new types of attractors in the system.

For bubbles ($R =1$) we find chaotic behaviour without memory with a
single symmetric attractor, as shown in a double periodic stroboscopic
representation in Fig.~\ref{fig:bub_2,0}a. It crosses the cell walls, i.e., the
movement while rising is not confined neither horizontally nor vertically. In
the horizontal motion of bubbles we find that there are two perceived time
scales: A fast local movement, and a slow cell-to-cell horizontal diffusion. We
can describe this as an intermittent horizontal spreading of trajectories
(Fig.~\ref{fig:diff_traj}a).

\begin{figure}[h]
  \begin{center}
    \includegraphics[width=.5\textwidth]{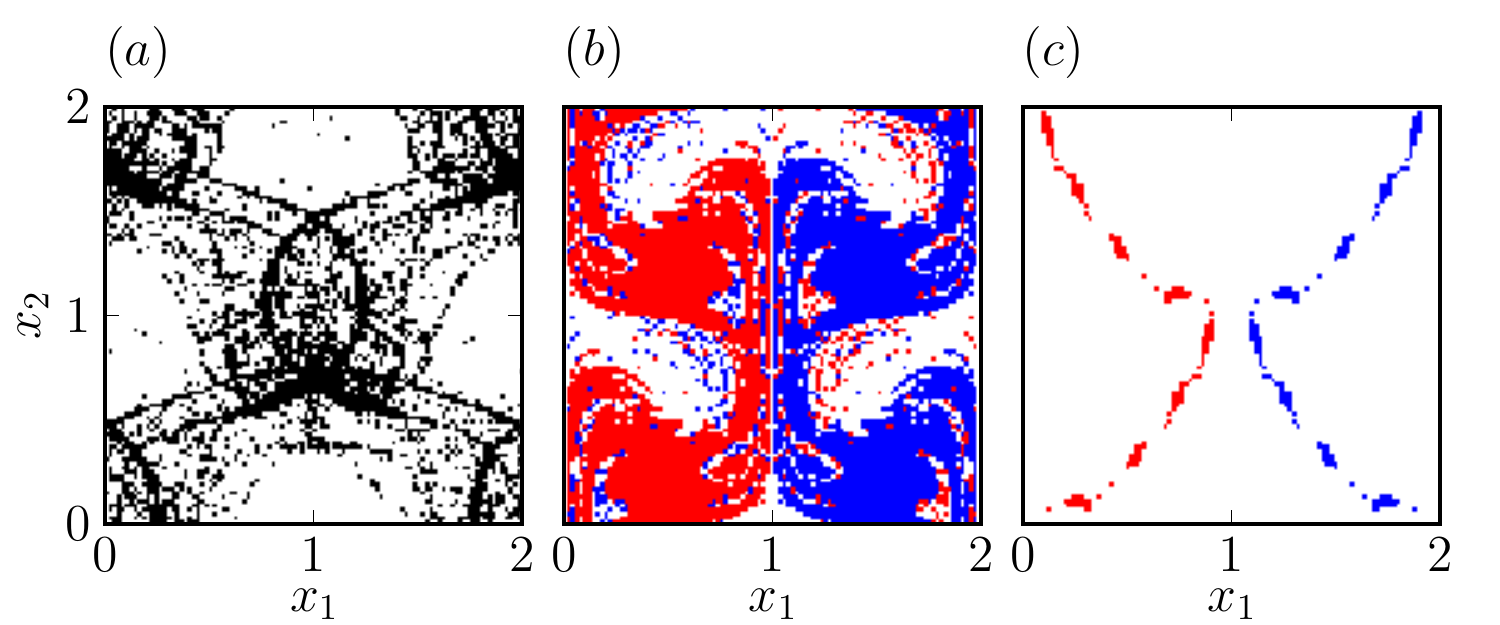}
    \end{center}
  \caption{Comparison of attractors and basins for bubbles with $\alpha=2,
    R=1$. (a) Single chaotic attractor without history force.  (b) Basins (white
    points represent initial conditions which have not reached the attractor in
    $120T$) and (c) two attractors for the dynamics with memory effects. All
    plots are obtained from the advection of $10^4$ particles at $t = 120T$ in
    the stroboscopic map.}\label{fig:bub_2,0}
\end{figure}

The inclusion of the history force has a very clear effect: It confines the
trajectories in the horizontal direction. This leads to the appearance of two
quasiperiodic attractors (Fig.~\ref{fig:bub_2,0}c) instead of a single chaotic
one (red curve in Fig.~\ref{fig:diff_traj}a).

The memoryless aerosol dynamics ($R = 0.5$) in this parameter range exhibits two
chaotic attractors, which are bounded in configuration space to the middle or to
the two vertical edges of the cell, respectively
(Fig.~\ref{fig:basin_2,2}b). The trajectories cross the vertical box borders
implying that sedimenting particles spread horizontally, but not further then
the next neighbor from which their return.
In the presence of the history force, the particles are no longer able to cross
the vertical cell borders, the motion becomes localized horizontally to a single
box width (Fig.~\ref{fig:basin_2,2}d).  In addition, instead of two
(Fig.~\ref{fig:basin_2,2}b), six attractors appear in 3 symmetric pairs, denoted
by I, I' and II in Fig.~\ref{fig:basin_2,2}d.  These attractors are no longer
chaotic, they exhibit quasi-periodic behaviour. One pair of the attractors (II)
meanders across the box, and the others (I, I') wiggle very close to the box
borders in the stroboscopic map, although they never cross the lines $x_1=0$ or
$1$ or $2$ (I and II in Fig.~\ref{fig:basin_2,2}; note the strong horizontal
magnification in I of the region near the box division). In (I) we see, when
plotting the force vectors again, that all point practically in the vertical
direction, and therefore there is no tendency for a border crossing due to
inertia. Note that there is a cusp in the attractor indicating a sign change in
the rotational direction of the vortices. In (II), the history force points
basically towards the vortex centers, helping to confine the trajectories to a
single box horizontally. The basins of attraction have fractal boundaries both
with and without memory (Fig.~\ref{fig:basin_2,2}a and c).
\begin{figure}[h]
  \begin{center}
    \includegraphics[width=.5\textwidth]{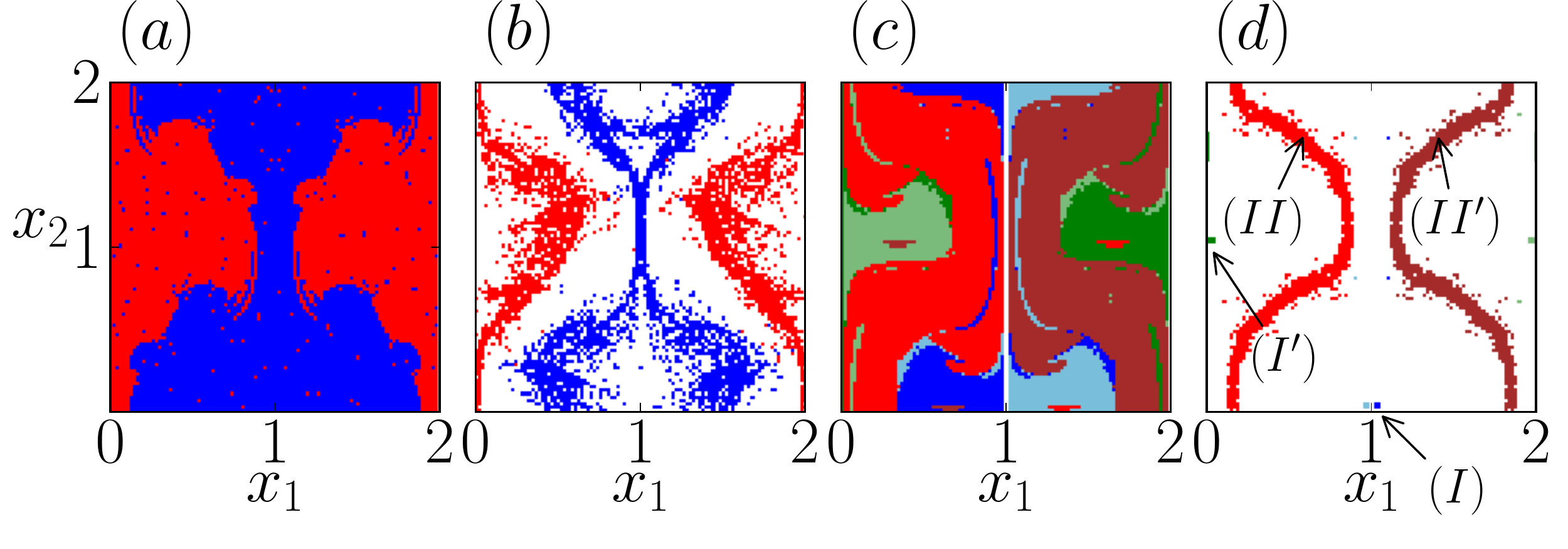}
    \includegraphics[width=.5\textwidth]{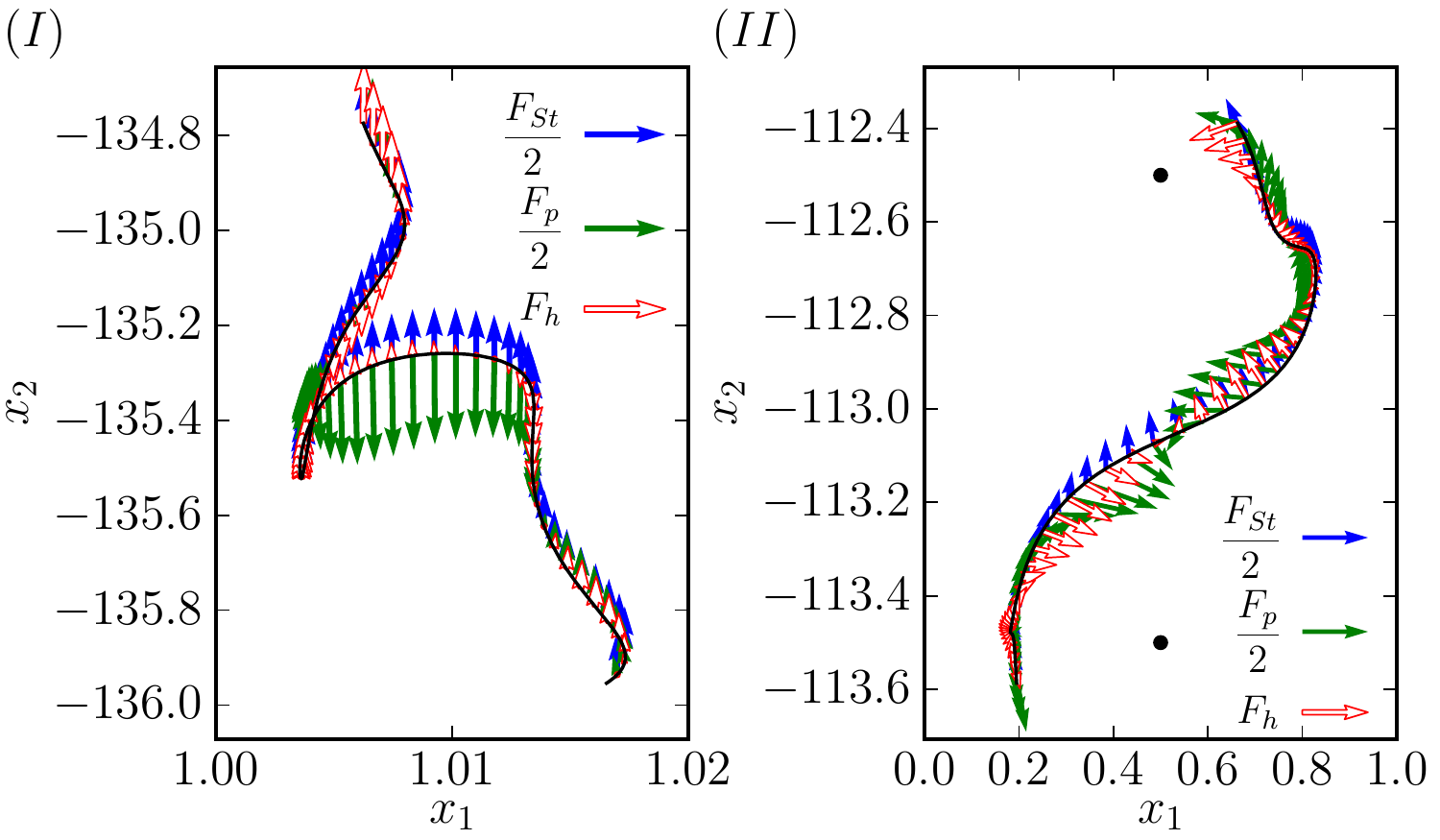}
    \end{center}
  \caption{Comparison for aerosols with $\alpha=2.2$. (a), (b): attractor
    structure without memory, basins (a) and attractors (b). The two chaotic
    attractors are marked by red and blue. (c), (d): attractor structure with
    memory, six attractors are present. The attractors are obtained from the
    advection of $10^4$ aerosol particles ($R = 0.5$) at $t = 120T$.
    (I), (II): trajectories on attractors I and II (plotted over $1.6T$, after
    discarding transients of $40T$) with the corresponding force
    vectors.}\label{fig:basin_2,2}
\end{figure}

\subsection{Changes in the nature of the attractor}

\begin{figure}[h]
  \begin{center}
    \includegraphics[width=.5\textwidth]{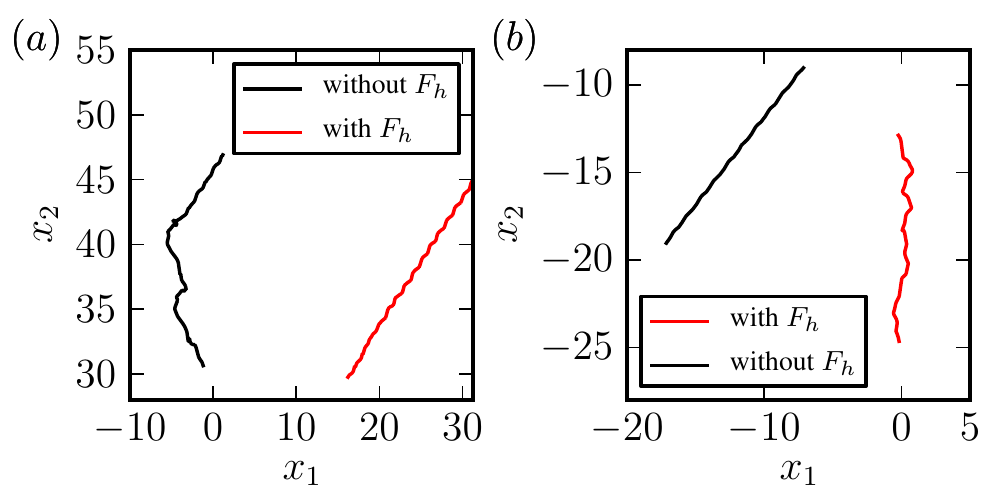}
    \end{center}
  \caption{ (a) Trajectories of bubbles $R=1$, $\alpha = 5.0$ over $30T$ on the
    attractor after $16T$ transients, in continues time representation. In the
    presence of history only one of the attractors is represented; (b)
    trajectories of aerosols $R=0.5$, $\alpha = 4.0$ on the attractor over $10T$
    after $10T$ transients, in continuous time representation. In the memoryless
    case only one of the attractors is
    represented.}\label{fig:traj_r1,0_alph5,0}\label{fig:halpha_diffusion}
\end{figure}

In the previous examples we already observed changes in the type of attractor
from periodic to quasiperiodic due to the impact of the history force. Besides
this transition, we find others which we discuss here with an example of a
conversion from periodic to chaotic dynamics, and vice versa.  The first occurs
for large aerosols, and the second for large bubbles.  In both situations the
magnitude of the history force is very strong, corresponding to $\sim 80\%$
(bubbles) and $\sim 35\%$ (aerosols) of the Stokes drag (see
Fig.~\ref{fig:forcas_dir}). As previously, there are strong changes of the
horizontal diffusivity of the advected particles. In the two examples the
periodic attractors are characterized by high diffusivity, where particles move
almost diagonally across the cells (Fig.~\ref{fig:halpha_diffusion}). We
characterise diffusivity by the variance of the displacements within a particle
ensemble. Since periodic attractors appear in symmetric pairs, this quantity
remains well defined in these cases. The chaotic attractors, on the other hand,
have trajectories with comparatively lower horizontal diffusion
(Fig.~\ref{fig:halpha_diffusion}). Here we again observe a change in the number
of attractors: We have a single one for the chaotic and two symmetric ones for
the periodic dynamics.

\subsubsection{From chaotic to periodic} 

The transition from a single chaotic attractor into two periodic ones we
exemplify with a case found for large bubbles $\alpha = 5.0$. The dynamics
without memory effects is chaotic:
The trajectories diffuse slowly horizontally from one cell to the next. The
single attractor in a double periodic representation shows a pronounced fractal
structure, Fig.~\ref{fig:basin_r1,0_alph5,0}a.  When the memory is included we
see two periodic attractors of period $2T$, which can be seen as two period-2
orbits in the double periodic representation of
Fig.~\ref{fig:basin_r1,0_alph5,0}c. The boundary of the basins of attraction for
these periodic attractors is of riddled type ~\cite{lai_transient_2011, alexander_riddled_1992},
that is the fractal dimension is nearly $2$
(Fig.~\ref{fig:basin_r1,0_alph5,0}b). As a consequence the transient times to
reach the attractors are very long ($\gg 100T$).

\begin{figure}[h]
  \begin{center}
    \includegraphics[width=.5\textwidth]{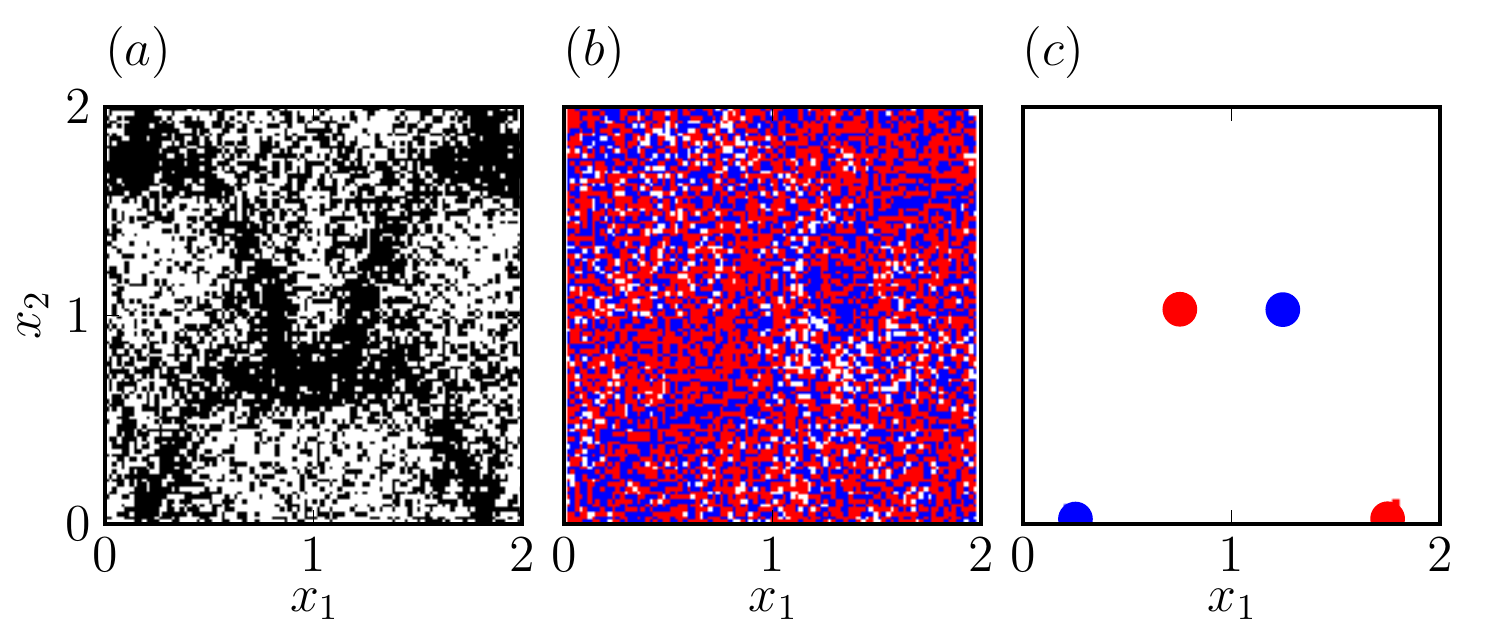}
    \end{center}
  \caption{Comparison for bubbles of $\alpha=5, R=1$. (a) Single chaotic
    attractor without memory.  (b) The basins of attraction and (c) two period-2
    attractors with memory.  All plots are obtained from the advection of $10^4$
    bubbles at $t = 120T$. White points in (b) represent initial conditions
    which have not reached the attractor in the given time
    interval.}\label{fig:basin_r1,0_alph5,0}
\end{figure}

These periodic attractors, in a continuous time representation, cross the cell
almost diagonally (Fig.~\ref{fig:traj_r1,0_alph5,0}c). The inclusion of the
history increases the diffusivity in the system, since the horizontal crossing
of the box boundaries is facilitated. To capture this effect we compute the
horizontal component of all the forces acting on the advected particle along the
trajectory as a function of $x_1$. In case of chaotic dynamics we work with
average values of forces felt by a single particle on the attractor in the
double periodic representation over a time of $500T$. For the periodic case we
just plot the horizontal component of the forces as a function of $x_1$. (We
consider forces as positive if they are directed from left to right.) We
observe that at the box boundaries among the horizontal component of all the
forces the history force provides the largest contribution, pulling the particle
from one box to the next (Fig.~\ref{fig:traj_r1,0_alph5,0}a). This results in an
enhanced horizontal diffusivity.

\begin{figure}[h]
  \begin{center}
    \includegraphics[width=.5\textwidth]{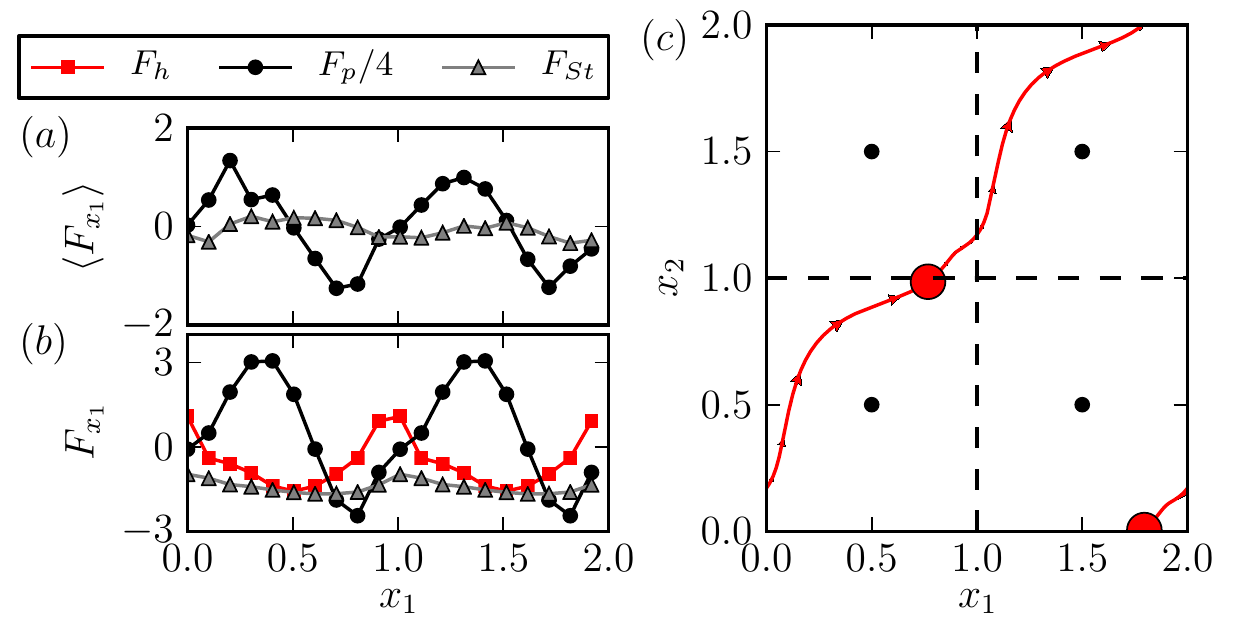}
    \end{center}
  \caption{Bubble dynamics for $\alpha=5, R=1$. Horizontal component of the
    forces at different positions $x_1$ computed along particle trajectory: (a)
    average forces on the chaotic attractor without memory; (b) forces on the
    attractor of panel (c). (c) Continuous time trajectory of the red attractor
    of Fig.~\ref{fig:basin_r1,0_alph5,0}c. Large red dots correspond to the
    positions of the attractor in the Poincare map. The centers of four vortices
    are represented as black dots.}\label{fig:traj_r1,0_alph5,0}
\end{figure}

\subsubsection{From periodic to chaotic}
We illustrate the transition from periodic to chaotic with a case of large
aerosols, $\alpha =4.0$. Without history force, we observe cell crossing
periodic trajectories spreading widely in the horizontal direction
(Fig.~\ref{fig:halpha_diffusion}b). The system exhibits two periodic attractors
of period $2T$, and basins of attraction with fractal boundaries
(Fig.~\ref{fig:basin_r0,5_alph4,0}a). When the history force is included the two
periodic attractors change into one chaotic attractor (see
Fig.~\ref{fig:basin_r0,5_alph4,0}b and c). The action of the history force is
opposite compared to the previous example: it weakens the strength of horizontal
diffusion.

\begin{figure}[h]
  \begin{center}
    \includegraphics[width=0.5\textwidth]{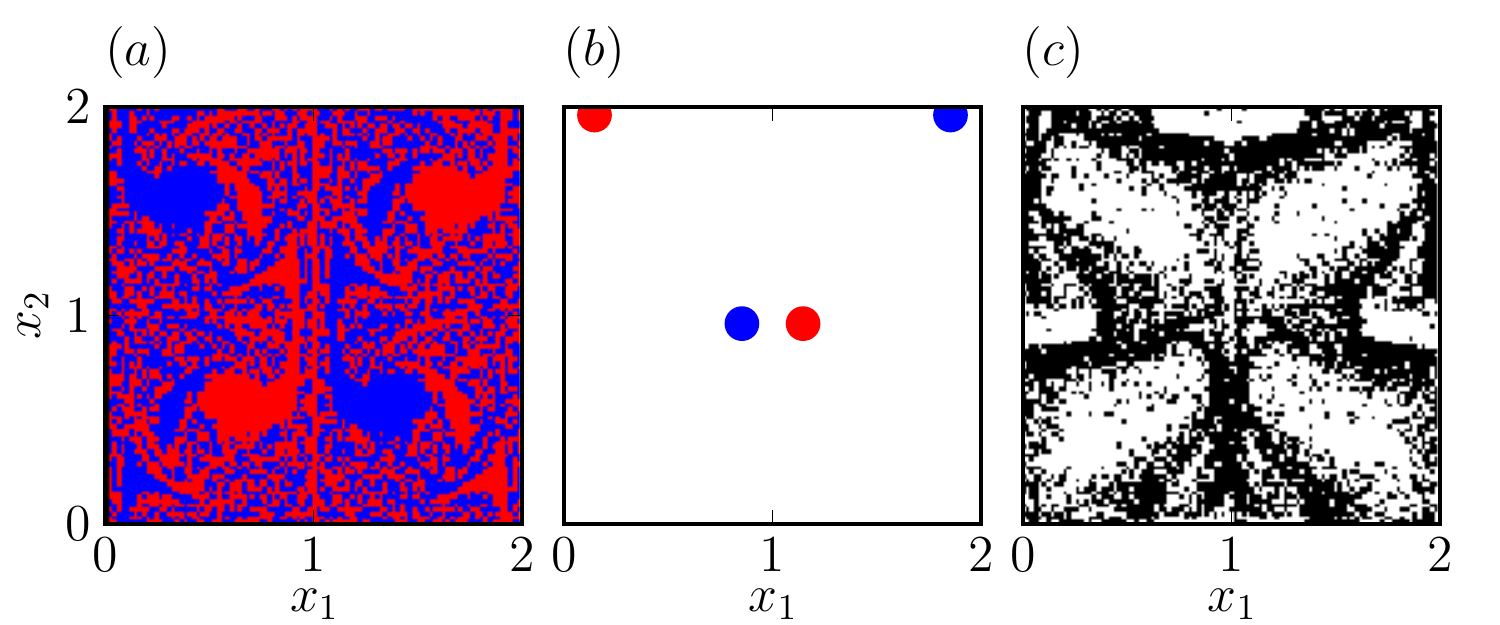}
    \end{center}
  \caption{Comparison for aerosols of $\alpha=4, R=0.5$.  The basins of
    attraction (a) and the corresponding periodic attractors (b) without
    memory. The chaotic attractor (c) in the presence of the history force. All
    plots determined from the advection of $10^4$ aerosol particles at
    $t=120T$.}\label{fig:basin_r0,5_alph4,0}
\end{figure}

To better understand the spreading behaviour we determine the horizontal
components of the three forces as a function of coordinate $x_1$. We use, as
previously, mean values over $500T$ for the dynamics on the chaotic
attractor. We observe that without memory, not only the pressure force controls
the horizontal movement, but also the Stokes drag (see
Fig.~\ref{fig:traj_r0,5_alph4,0}a). The graph of the pressure force is
asymmetric and the positive part dominates, the Stokes drag has permanently
negative sign. (The signs of the forces are the opposite on the symmetric
counterpart of this attractor.)
When the memory is included, we observe a complete change of the horizontal
components of the forces (see Fig.~\ref{fig:traj_r0,5_alph4,0}b); the forces
have no longer a preferential direction, and decrease in magnitude. The
horizontal components of the three forces provide a kind of string force
pointing towards the cell center. In the presence of the history force the
positive and negative contributions of the horizontal forces nearly balance each
other. As a consequence it is easier for the trajectories to turn, which results
in an increase of the curvature. The motion withing a cell is followed also in
this case by a cell-to-cell horizontal displacement. We can describe this again
as an intermittent horizontal spreading of trajectories
(Fig.~\ref{fig:basin_r0,5_alph4,0}c).


\begin{figure}[h]
   \includegraphics[width=.5\textwidth]{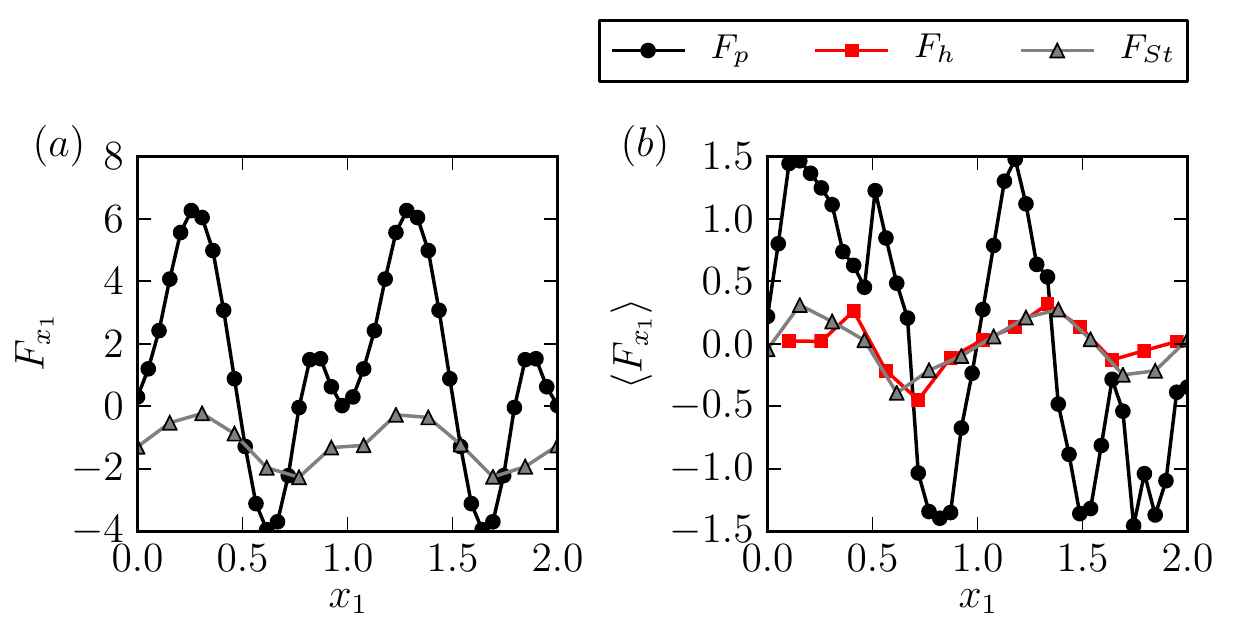}
  \caption{ Horizontal component of the forces acting on an aerosol particle as
    a function of $x_1$: (a) computed for the blue attractor without history
    force (Fig.~\ref{fig:basin_r0,5_alph4,0}b); (b) with history force. Here, as
    the trajectory is chaotic, we plot a temporal average over
    $500T$.}\label{fig:traj_r0,5_alph4,0}
\end{figure}

Having described the main effects of the history force on the attractors
of the system, we now move to the effect on the transport properties of
the flow.

\section{Transport properties}\label{sec:transport}

\subsection{Horizontal spreading}
\begin{figure}[h!]
  \begin{center}

    \includegraphics[width=.23\textwidth]{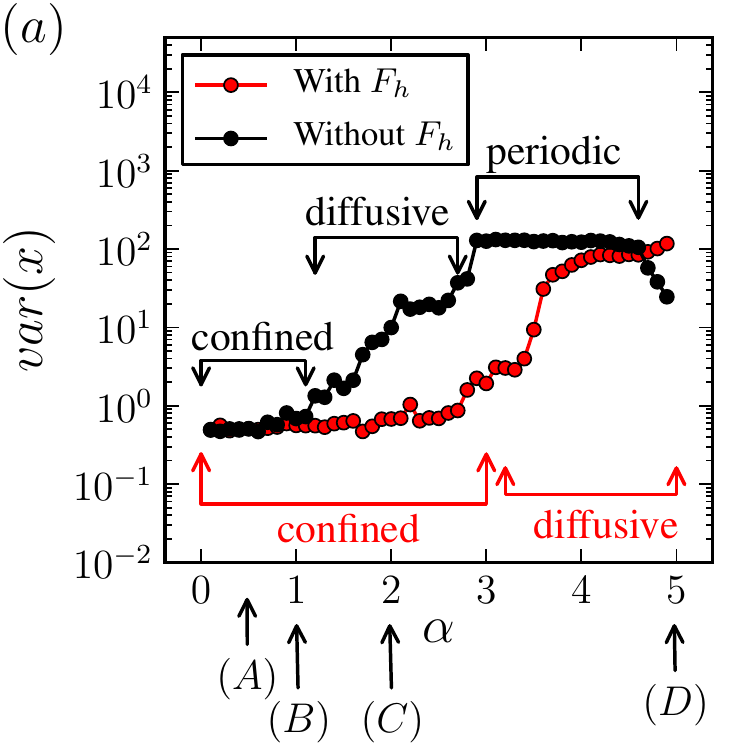}
    \includegraphics[width=.23\textwidth]{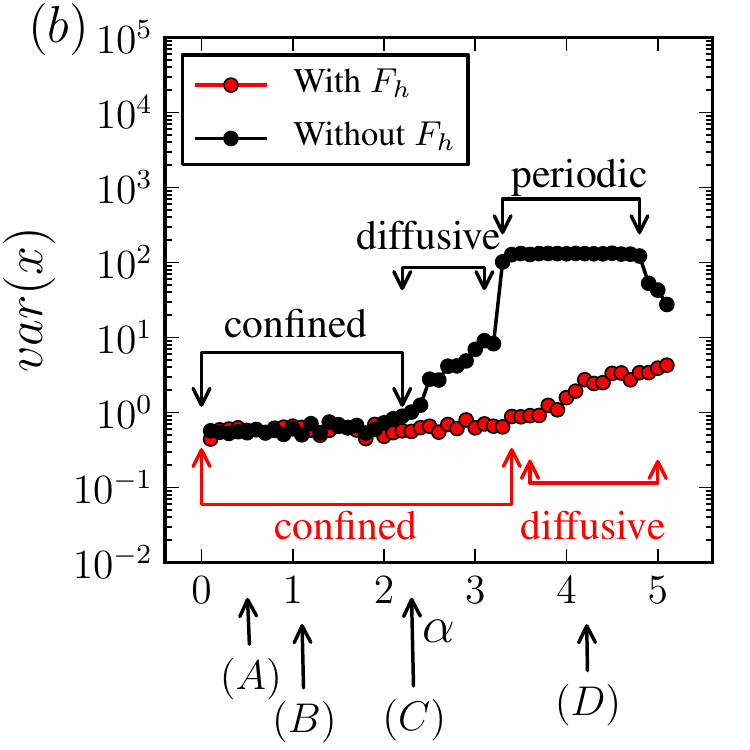}
    \end{center}
  \caption{Variance of horizontal displacement with and without memory,
    evaluated at $t = 140T$.  
    $\alpha$ values corresponding to cases treated in
    Sec.~\ref{sec:bifurcation} are marked by labels $A, \dots,
    D$}\label{fig:diff}
\end{figure}

Here we compare the transport properties in the horizontal direction with and
without the history force. We observe that the history force can introduce
strong changes in the horizontal dynamics: in a wide range of parameters, for
both aerosols and bubbles, the effect of memory results in weakening or even
blocking the spreading. To quantify this effect, we determine the horizontal
displacements $x_{1,t}$ from the initial position for $N=30$ trajectories after
a fixed time interval $t=140T$.  The initial conditions are randomly chosen
within a cell.  We evaluate the mean $\left<x_t\right>$ and the variance
\begin{equation}
\mbox{var} (x) =\left( \left< x^2_t\right>-\left<x_t\right>^2 \right)^{1/2}
\end{equation}
over the ensemble.  The results for the variance are plotted in
Fig.~\ref{fig:diff}a and b.  We note that the computational demands do not allow
us to evaluate the variance over longer time intervals (and larger ensembles)
from which diffusion coefficients could be determined with some confidence.  In
addition, the intermittent behavior mentioned earlier between staying within a
cell and cross-cell horizontal motion makes us believe that in some cases the
spreading is superdiffusive. With these constraints we characterize horizontal
diffusivity in terms of the variance. Note that because of using several initial
conditions, the behavior over all attractors is averaged.

As a function of the size parameter, we can distinguish
three regimes in Fig.~\ref{fig:diff}a and b.
Particles of small sizes 
stay bound to the cell
where they are initiated, and the variance is below unity. This region is marked as
``confined''.
For larger particles, we see a tendency of increase (beyond unity) in the
variance due to spreading into the neighboring cells. The strength of spreading
increases with the particle size. This interval of $\alpha$ is called
``diffusive''.  For even larger particles the movement on the attractors turns
into a ballistic-like motion where particle cross the cells almost
diagonally. This periodic motion corresponds to the approximately horizontal
plateaus in Fig.~\ref{fig:diff}, which we call ``periodic''.

With the history force, the horizontal confinement to the original cell persists
through a wider range of sizes. Diffusion appears at about $\alpha = 2.8$ for
bubbles and at about $\alpha = 3.5$ for aerosols, Fig.~\ref{fig:diff}. This
horizontal confinement up to $\alpha = 2.8$ is in harmony with the lower panels
of the bifurcation diagrams of Fig.~\ref{bifurcation}. In the interval $2.8 <
\alpha < 3.5$ for aerosols the stretching of the bifurcation diagram over the
unit interval only implies a communication between neighbouring cells without
further spreading.

\begin{figure}[h!]
  \begin{center} 
    \includegraphics[width=.23\textwidth]{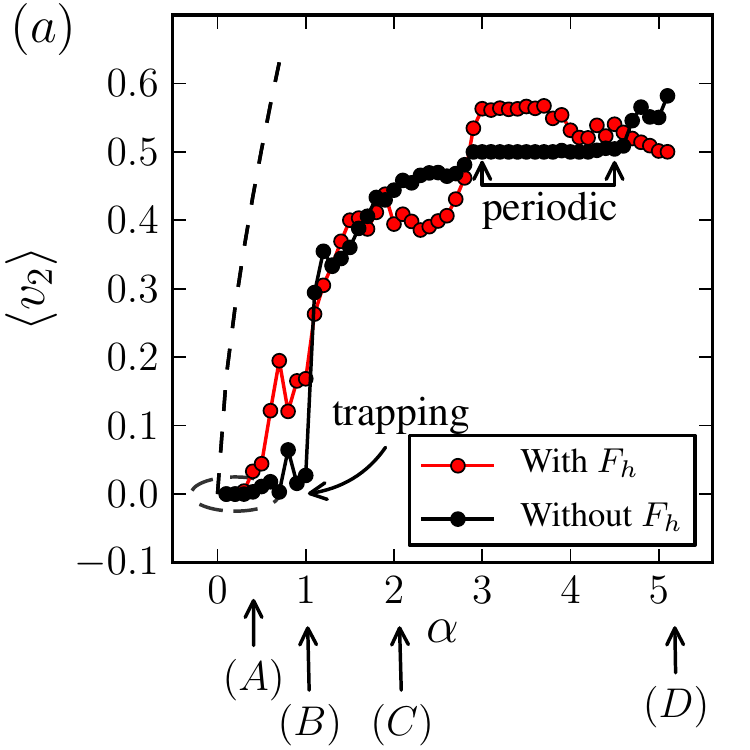}
    \includegraphics[width=.23\textwidth]{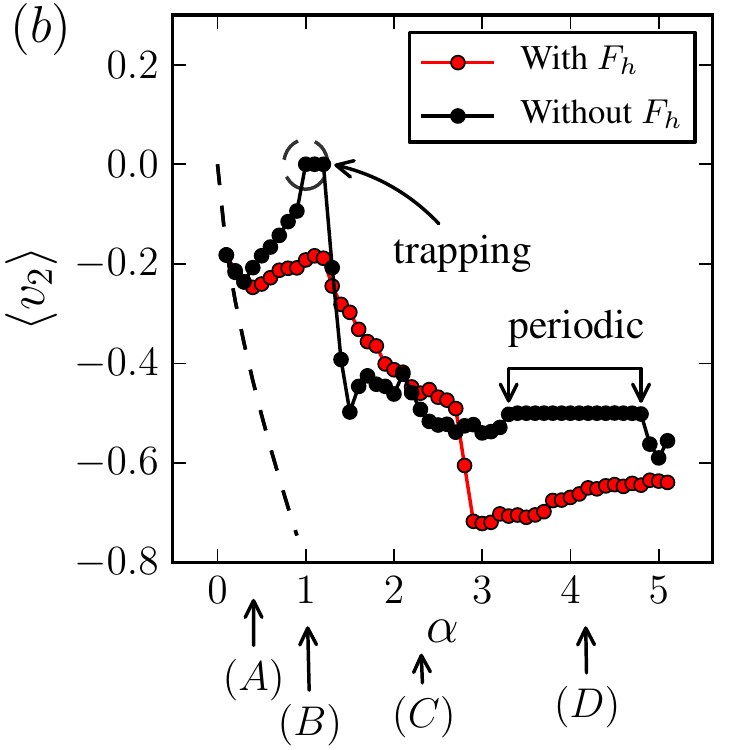}
  \end{center}
  \caption{Comparison of average vertical velocities $\left<v_{2}\right>$ with
    and without memory. (a): Bubbles, $R = 1.0$; (b): Aerosols, $R = 0.5$.  The
    $\alpha$ values corresponding to cases treated in
    Sec.~\ref{sec:bifurcation} are marked by labels $A, \dots, D$. The
    settling/ raising velocities in a fluid at rest are plotted with dashed
    lines.}
  \label{fig:vel_vert}
\end{figure}


\subsection{Vertical velocity}
Another important feature is transport in the vertical direction.  A useful
quantity for characterizing its strength is the average vertical velocity,
obtained again over an ensemble of trajectories. The average values are
computed for 30 randomly initiated trajectories within a cell over $50T$ after
cancelling $50T$ transients. This quantity is also affected by the history
force, the most striking effect being the disappearance of vertical trapping
($\left<v_{2}\right>=0$) for aerosols of any size (see Fig.~\ref{fig:vel_vert}b).  For
bubbles the history force reduces the range of trapping, which is only possible
for very small bubbles, up to $\alpha = 0.4$ (in a size parameter range about
$2$ times smaller than without memory), as can be seen in
Fig.~\ref{fig:vel_vert}a 
The action of the forces, in the given flow, determines the shape of the
attractors in a given phase instance. The average velocity on the attractors
depends on their shape, and this might result e.g. in lower velocities when
larger portions of the attractor are close to vortex centers where the flow is
weak.  This explains the larger sinking velocities in the presence of memory for
$\alpha>3$ than those without. Periodic trajectories belong again to the
plateaus of velocities. It is interesting to compare these results with those
valid in a fluid at rest. The dimensionless vertical velocity is given then by
Eqs.~\ref{W} and \ref{w}. In view of Eq.~\ref{w1}, and the fact that $w_1 = 1.6$
we find for our bubbles and aerosols that $W = \pm 0.8 \alpha^{2/3}$. These
curves are plotted as dashed lines in Fig~\ref{fig:vel_vert}, they correspond to
much larger velocities than the measured ones. This difference shows how
important the effect of the flow is on the vertical motion, $W$ cannot even be
used as an estimate.

\section{Conclusion}\label{sec:conclusion}

We compared the dynamics of chaotic advection of the finite size particles using
the full Maxey-Riley equations with the well spread approximation which ignores
the history force term. We observed strong differences between both dynamics,
such as a large increase in the transient time, and a modification of the nature
and number of attractors. Directly from the equations, one can see that the
effect of the history force increases with the particle size.  For small
particles, the history force is relatively weak compared to other forces, it is,
however, already large enough to affect the shape and the position of the
attractors in configuration space, and in particular it causes the basins of
attraction to acquire fractal boundaries. For slightly larger particles, for
which particle trapping in a single box is possible in the memoryless case, the
history force generates vertical leaking so that the resulting trajectories jump
from one cell to the next in the vertical. The fact that trapping fully
disappears for aerosols, and becomes restricted to very small bubbles only, is
one of the most striking effects of memory.  This might be associated with a
decrease in the number of attractors. With a further increase of the size
parameter, the history force strongly decreases the horizontal spreading of
particles, it might even prevent the attractors to cross from one box to the
next in the horizontal. This weakening in horizontal spreading can go along with
an increases in the number of attractors. We also find that the effect of
history can be the opposite: in such cases the history force pulls the particles
to the next box, resulting in a change in the type of attractor from chaotic to
periodic.  This might be the dynamical background of the changes found in the
transport properties of the system.

Based on the strong differences found with the inclusion of memory
effects, we argue that it is essential to consider the Maxey-Riley
equations in their full form, in order to account for the full complexity of
advection dynamics.

The impact of the history force on transport properties of inertial
  particles in flows may have significant implications for sedimenting particles
  in the ocean. Vertical export of marine aggregates is considered to be one
  important process to export carbon into deeper ocean layers~\cite{
    de_la_rocha_factors_2007}. Here often settling velocities (Eq.~\ref{W}) are
  used to estimate fluxes of particles. However, our results show, that these
  settling velocities (dashed line in Fig.~\ref{fig:vel_vert}) turn out to be
  much faster than the ones obtained from moving inertial particle in a
  flow. Therefore, export rates of aggregates based on settling velocities in
  still fluids appear to be dramatically overestimated compared to the ones which
  take the motion in a fluid flow explicitly into account. 


\section{Acknowledgment}
We would like to thank A. Daitche for valuable discussions. We acknowledge
support from COST-Action MP0806 ``Particles in Turbulence''. U.F. would like to
thank R. Roy and his group for their hospitality and acknowledges support from
the ``Burgers Program for Fluid Dynamics'' of the University of Maryland at
College Park. T.T. acknowledges the support of OTKA grant NK100296 and of the
Alexander von Humboldt Foundation.

\bibliographystyle{apsrev4-1}
\bibliography{bib}

\end{document}